\documentclass{article}

\usepackage{microtype}
\usepackage{graphicx}
\usepackage{booktabs} %

\usepackage[breaklinks]{hyperref}

\usepackage[accepted]{icml2023}

\usepackage{amsmath}
\usepackage{amssymb}
\usepackage{mathtools}
\usepackage{amsthm}

\usepackage{url}
\usepackage[utf8]{inputenc} %
\usepackage[T1]{fontenc}    %
\usepackage[breaklinks]{hyperref}      %
\usepackage{url}            %

\usepackage{booktabs}       %
\usepackage{amsfonts}       %
\usepackage{amsmath}
\usepackage{amssymb}
\usepackage{mathtools}
\usepackage{amsthm}
\usepackage{nicefrac}       %
\usepackage{microtype}      %
\usepackage{xcolor}         %
\usepackage{wrapfig}
\usepackage{graphicx}
\usepackage{caption}
\usepackage{multirow,makecell}

\usepackage{xspace}
\usepackage{subcaption}

\usepackage{algorithm}
\usepackage{algorithmic}

\usepackage[capitalize,noabbrev]{cleveref}

\theoremstyle{plain}

\theoremstyle{definition}

\theoremstyle{remark}

\usepackage[textsize=tiny]{todonotes}

\icmltitlerunning{SWARM Parallelism: Training Large Models Can Be Surprisingly Communication-Efficient}

\begin{document}

\twocolumn[
\icmltitle{SWARM Parallelism: Training Large Models\\ Can Be Surprisingly Communication-Efficient}

\icmlsetsymbol{equal}{*}

\begin{icmlauthorlist}
\icmlauthor{Max Ryabinin}{equal,hse,yandex}
\icmlauthor{Tim Dettmers}{equal,uw}
\icmlauthor{Michael Diskin}{yandex,hse}
\icmlauthor{Alexander Borzunov}{hse,yandex}
\end{icmlauthorlist}

\icmlaffiliation{hse}{HSE University}
\icmlaffiliation{yandex}{Yandex}
\icmlaffiliation{uw}{University of Washington}

\icmlcorrespondingauthor{Max Ryabinin}{mryabinin0@gmail.com}

\icmlkeywords{Machine Learning, ICML}

\vskip 0.3in
]

\printAffiliationsAndNotice{\icmlEqualContribution} %

\begin{abstract}

Many deep learning applications benefit from using large models with billions of parameters. Training these models is notoriously expensive due to the need for specialized HPC clusters. In this work, we consider alternative setups for training large models: using cheap ``preemptible'' instances or pooling existing resources from multiple regions. We analyze the performance of existing model-parallel algorithms in these conditions and find configurations where \textit{training larger models becomes less communication-intensive}.
Based on these findings, we propose SWARM parallelism\footnote{SWARM parallelism is a backronym for Stochastically Wired Adaptively Rebalanced Model Parallelism.}, a model-parallel training algorithm designed for poorly connected, heterogeneous and unreliable devices. SWARM creates temporary randomized pipelines between nodes that are rebalanced in case of failure. 
We empirically validate our findings and compare SWARM parallelism with existing large-scale training approaches.
Finally, we combine our insights with compression strategies to train a large Transformer language model with 1B shared parameters (${\approx}13$B before sharing) on preemptible T4 GPUs with less than 200Mb/s network.

\end{abstract}

\section{Introduction}
\label{sect:intro}

For the past several years, the deep learning community has been growing more reliant on large pretrained neural networks. The most evident example of this trend is natural language processing, where the parameter count of models has grown from hundreds of millions~\citep{transformer,gpt,bert} to billions~\citep{megatron2,t5,gptj,ernie3} to hundreds of billions~\citep{gpt3,fedus2021switch,palm,gopher} with consistent gains in quality~\citep{kaplan2020scaling}. Likewise, many models in computer vision are reaching the billion-parameter scale~\citep{dalle,scaling_vit,coatnet,guided_diffusion}.

At this scale, the models no longer fit into a single accelerator and require specialized training algorithms that partition the parameters across devices~\citep{alexnet,dean12}. While these model-parallel algorithms use different partitioning strategies, they all share the need to perform intensive device-to-device communication~\citep{pipedream,megatron2}. Also, if a single device fails, it will cause the entire training process to break down. As a result, model-parallel algorithms are typically deployed in dedicated high-performance computing (HPC) clusters or supercomputers~\citep{shoeybi2019megatron,zero,megatron2}.

This kind of infrastructure is notoriously expensive to build and operate, which makes it available only to a few well-resourced organizations~\citep{summit,fugaku,microsoft_supercomputer}. Most researchers cannot afford the experiments necessary for a proper evaluation of their ideas. This ultimately limits the scientific progress for many important research areas, such as solving NLP problems in ``non-mainstream'' languages.

Several recent works propose more cost-efficient distributed training strategies that leverage fleets of temporary ``preemptible'' instances that can be dynamically allocated in regions with low demand for hardware and electricity, making them 2--10 times cheaper than their dedicated counterparts~\citep{proteus}. Another solution is to train in ``collaborations'' by pooling together preexisting resources or using the help of volunteers~\citep{dedloc,eydle,hivemind_dmoe,yuan2022decentralized}. 

However, training in either of those setups requires specialized algorithms that can adapt to the changing number of workers, utilize heterogeneous devices and recover from hardware and network failures. While there are several practical algorithms for unreliable hardware~\citep{volunteer_dl_async,lin2020multinode,moshpit}, they can only train relatively small models that \textit{fit into the memory of the smallest device}. This limits the practical impact of cost-efficient strategies, because today's large-scale experiments often involve models with billions of parameters.%

In this work, we aim to find a practical way of training large neural networks using \textbf{unreliable heterogeneous devices with slow interconnect}.
We begin by studying the impact of model size on the balance between communication and computation costs of pipeline-parallel training.
Specifically, increasing the size leads computation costs to grow faster than the network footprint, thus making \textbf{household-grade connection speeds} more practical than one might think.
This idea inspires the creation of \textbf{SWARM parallelism}, a pipeline-parallel approach designed to handle peer failures by prioritizing stable peers with lower latency.
In addition, this approach periodically rebalances the pipeline stages, which allows handling devices with different hardware and network speeds.

In summary, we make the following contributions:
\vspace{-6pt}

\begin{itemize}
    \item We analyze the existing model-parallel training techniques and formulate the ``Square-Cube Law'' of distributed training: a counterintuitive observation that, for some methods, \textit{training larger models can actually decrease the network overhead}.
    \item We develop SWARM parallelism, a decentralized model-parallel algorithm\footnote{The code for our experiments can be found at \href{https://github.com/yandex-research/swarm}{\texttt{github.com/yandex-research/swarm}}.}that leverages randomized fault-tolerant pipelines and dynamically rebalances nodes between pipeline stages. To the best of our knowledge, this is the first decentralized algorithm capable of billion-scale training on heterogeneous unreliable devices with slow interconnect.
    \item Combining insights from the square-cube law, SWARM parallelism, and 8-bit compression, we show that it is possible to train a billion-scale Transformer language model on preemptible servers with low-power GPUs and the network bandwidth of less than $200$Mb/s while achieving high training throughput.
\end{itemize}

\section{Background \& Related Work}
\subsection{Model-Parallel Training}\label{sect:related_model_parallel}

Over the past decade, the deep learning community has developed several algorithms for training large neural networks. 
Most of them work by dividing the model between multiple workers, which is known as model parallelism. 
The exact way in which these algorithms divide the model determines their training performance and the maximum model size they can support.

\paragraph{Traditional model parallelism.} Historically, the first general strategy for training large models was to assign each device to compute a subset of each layer (e.g., a subset of neurons), then communicate the results between each other~\citep{alexnet,model_parallelism_survey1,model_parallelism_survey2}.
Since each device stores a fraction of layer parameters, this technique can train models with extremely wide layers that would not fit into a single GPU. However, applying traditional model parallelism to deep neural networks comes at a significant performance penalty, as it requires all-to-all communication after each layer.
As a result, while intra-layer parallelism is still widely used~\citep{meshtensorflow,zero}, it is usually applied within one physical server in combination with other strategies~\citep{krizhevsky2014oneweirdtrick,projectadam,beyond_data_and_model,megatron2}.

\vspace{-8pt}
\paragraph{Pipeline parallelism} circumvents the need for expensive all-to-all communication by assigning each device with one or several layers~\citep{huang2019gpipe}. During the forward pass, each stage applies its subset of layers to the inputs supplied by the previous stage, then sends the outputs of the last layer to the next stage. For the backward pass, this process is reversed, with each pipeline stage passing the gradients to the device that supplied it with input activations.

To better utilize the available devices, the pipeline must process multiple microbatches per step, allowing each stage to run in parallel on a different batch of inputs. In practice, the number of microbatches is limited by the device memory: this results in reduced device utilization when processing the first and the last microbatches, known as the ``bubble'' overhead~\citep{huang2019gpipe}. To combat this issue, subsequent studies propose using activation checkpointing, interleaved scheduling, and even asynchronous training~\citep{pipedream,megatron2,huang2019gpipe,shoeybi2019megatron,pipemare}.

Aside from model parallelism, there two more strategies for training large models: data parallelism with dynamic parameter loading~\citep{zero} and model-specific algorithms such as Mixture-of-Experts~\citep{shazeer2017outrageously}. We discuss these algorithms in Appendix~\ref{appendix:related} and compare the performance of offloading with SWARM in Section~\ref{appendix:training_throughput} and Appendix~\ref{appendix:equivalence}.

\vspace{-6pt}
\subsection{Distributed Training Outside HPC}
\label{sect:related_cost_efficent_collaborative}
\vspace{-2pt}

The techniques described in Section~\ref{sect:related_model_parallel} are designed for clusters of identical devices with rapid and reliable communication, making them a natural fit for the HPC setup. As we discussed earlier, such infrastructure is not always available, and a more cost-efficient alternative is to use ``preemptible'' instances~\citep{li2019speeding,zhang2020machine,proteus} or volunteer computing~\citep{volunteer_dl_async,hivemind_dmoe,eydle,dedloc}. However, these environments are more difficult for distributed training: each machine can disconnect abruptly due to a failure or preemption. Besides, since there is a limited number of available instances per region, training at scale often requires operating across multiple locations or using different instance types.

To handle unstable peers and heterogeneous devices, the research community has proposed elastic and asynchronous training methods, correspondingly.
Moreover, training large models over heterogeneous devices can be optimized with global scheduling~\cite{yuan2022decentralized}.
We describe these methods in more detail in Appendix~\ref{appendix:related}; importantly, neither of them are unable to satisfy all the constraints of our setup.

By contrast, the largest models have billions of parameters, which exceeds the memory limits of most low-end computers. 
However, model-parallel algorithms are not redundant, which makes them more vulnerable to hardware and network failures. 
There exist two methods that allow training large models with unreliable devices~\citep{hivemind_dmoe,thorpe2022bamboo}: however, the first one supports only specific architectures and requires at least 1Gb/s bandwidth, whereas the second one has no publicly available implementations, relies on redundant computations for fault tolerance and considers only the homogeneous setup.

\vspace{-6pt}
\subsection{Communication Efficiency and Compression}~\label{sect:related_communication_eficiency}
\vspace{-18pt}

In this section, we discuss techniques that address training with limited network bandwidth or high latency, such as gradient compression or overlapping computation with communication phases. These techniques are often necessary for distributed training without high-speed connectivity, because otherwise the performance of the system becomes severely bottlenecked by communication.

\vspace{-6pt}
\paragraph{Efficient gradient communication.}

Data-parallel training requires synchronization of gradients after each backward pass, which can be costly if the model has many parameters or the network bandwidth is limited. There exist several methods that approach this problem: for example, Deep Gradient Compression~\citep{deepgradientcompression} sparsifies the gradients and corrects the momentum after synchronization, while PowerSGD~\citep{vogels2019powersgd} factorizes the gradients and uses error feedback to reduce the approximation error.
Recently, \citet{wang2022finetuning} proposed to compress the changes of model activations, achieving high-speed communication for finetuning models of up to 1.5B parameters.
Alternatively, \citet{Dettmers20158BitAF} uses 8-bit quantization to compress gradients before communication. We evaluate it along with compression-aware architectures, leaving the exploration of more advanced approaches to future work.

Besides gradient compression, another effective technique is to use layer sharing~\citep{albert}, which reduces the number of aggregated gradients by a factor of how many times each layer is reused.

\paragraph{Overlapping communication and computation.}

Model, pipeline, and data parallelism all have synchronization points and require transfer of gradients or activations. One way to reduce the transfer cost is to overlap communication with computation, {\it hiding} the synchronization latency. This overlap can be achieved by combining parallelism techniques~\citep{krizhevsky2014oneweirdtrick, zero}, by synchronizing gradients layer-by-layer in lockstep with backpropagation~\citep{paszke2019pytorch}, or by using pure pipeline parallelism~\citep{huang2019gpipe,pipedream}.
However, pure pipeline parallelism requires many stages to effectively hide the latency. To overcome this problem, we study inter-layer compression techniques that work well even with relatively few pipeline stages.

\begin{figure*}[t]
    \centering
    \begin{minipage}[][][b]{0.64\textwidth}
    \includegraphics[width=\linewidth]{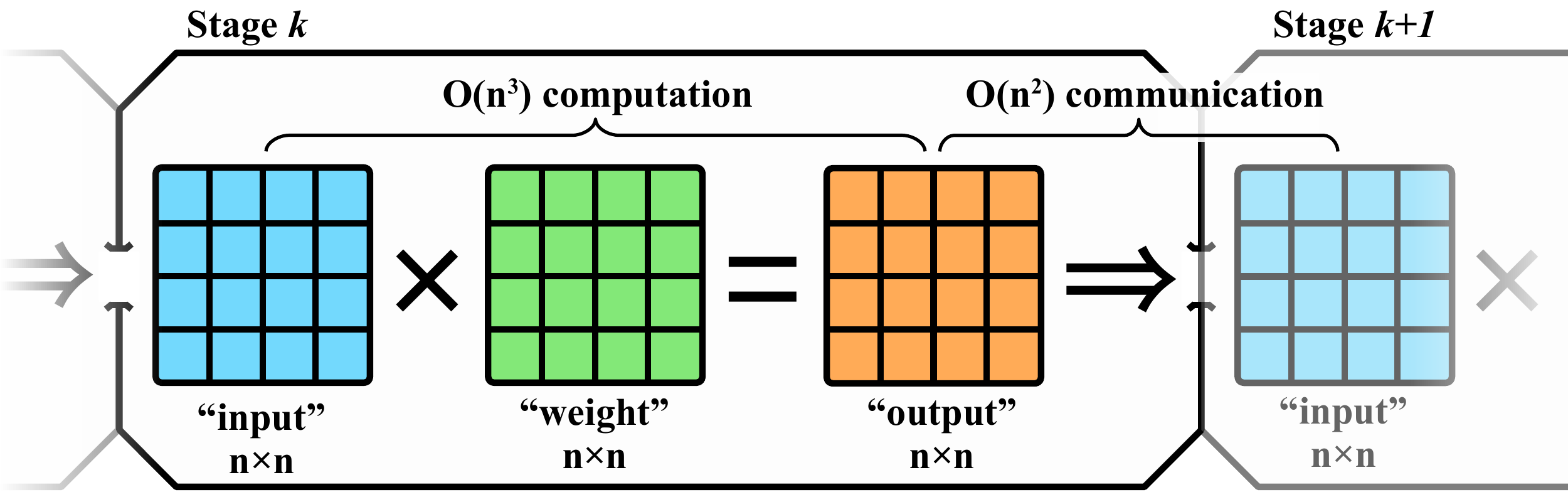}
    \caption{\textbf{(Left)} An intuitive explanation of the square-cube law, \textbf{(Right)} Relative device utilization for Transformer layers using Tesla V100 and 500Mb/s network bandwidth. See Section~\ref{sect:experiments_square_cube} and Appendix~\ref{appendix:detailed_setup} for a detailed setup.}
    \label{fig:squarecube}
    \end{minipage}
    \begin{minipage}[][][b]{0.35\textwidth}
    \includegraphics[width=\linewidth]{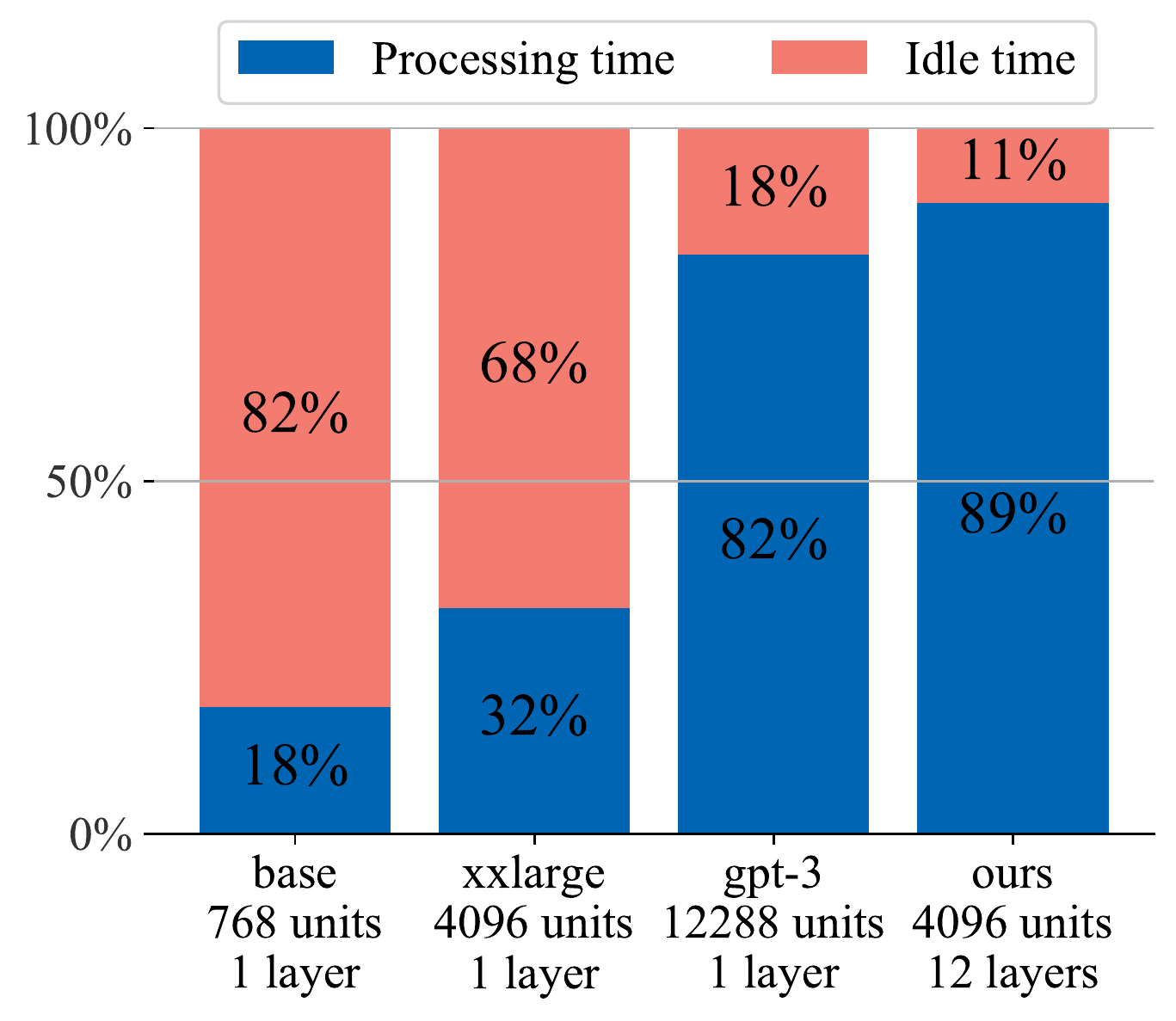}
    \end{minipage}
    \vspace{-10pt}
\end{figure*}

\section{Communication-Efficient Model Parallelism}\label{sect:method}
\vspace{-4pt}

In this section, we outline our approach for training large models with heterogeneous unreliable poorly-connected devices.
To that end, the section is organized as follows:

\vspace{-4pt}
\begin{itemize}
    \item Section~\ref{sect:method_squarecube} analyzes how existing model-parallel algorithms scale with model size and shows conditions where training increasingly larger models leads to less intense network usage;
    \item Section~\ref{sect:method_swarm} describes SWARM parallelism --- a decentralized algorithm for training large models under the conditions outlined in Section~\ref{sect:related_cost_efficent_collaborative}.
\end{itemize}
\vspace{-12pt}

\subsection{The Square-Cube Law of Distributed Training}\label{sect:method_squarecube}

To better understand the general scaling properties of model parallelism, we need to abstract away from the application-specific parameters, such as model architecture, batch size, and system design. To that end, we first consider a simplified model of pipeline parallelism. Our ``pipeline'' consists of $k$ stages, each represented by $n{\times}n$ matrices. Intuitively, the first matrix represents the input data and all subsequent matrices are linear ``layers'' applied to that data. This model abstracts away from application-specific details, allowing us to capture general relationships that hold for many models.

During ``training'', stages iteratively perform matrix multiplication and then send the output to the subsequent pipeline stage over a throughput-limited network. These two operations have different scaling properties.
The compute time for naïve matrix multiplication scales as $O(n^3)$. While this can be reduced further in theory~\citep{coppersmith_winograd,refined_laser}, it is only used for very large matrices~\citep{practical_matmul_best,practical_matmul_earlier,strassen_reloaded}. Therefore, deep learning on GPUs typically relies on $O(n^3)$ algorithms.

In turn, the communication phase requires at most $O(n^2)$ time to transfer a batch of $n{\times}n$ activations or gradients. Therefore, as we increase the model size, the computation time grows faster than communication time, regardless of which matrix multiplication algorithm we use. We refer to this idea as the \textit{square-cube law} after the eponymous principle in physics~\citep{square_cube,mechanics}.

This principle applies to many real-world neural network architectures, albeit with some confounding variables. In convolutional neural networks~\cite{conv_first}, the computation time scales as $O(B H W C^2)$ and the communication is $O(B H W C)$, where $B$, $H$, $W$ and $C$ stand for batch size, height, width and the number of channels. Recurrent neural networks~\citep{backprop_rnn,lstm} need $O(B L H^2)$ compute in terms of batch size, sequence length, and hidden size, respectively, and $O(B L H)$ or $O(B H)$ communication, depending on the architecture. With the same notation, Transformers~\citep{transformer} require $O(B L^2 H)$ compute for attention layers, $O(B L H^2)$ compute for feedforward layers, but only $O(B L H)$ communication.%

Based on these observations, we conclude that pipeline parallelism naturally grows more communication-efficient with model size. More precisely, increasing the hidden dimension will reduce the communication load per device per unit of time, making it possible to train the model efficiently \textit{with lower network bandwidth} and \textit{higher latency}\footnote{Latency slows the communication down by a constant factor that also grows less important with model size.}. While the exact practical ramifications depend on the use case, Section~\ref{sect:experiments_square_cube} demonstrates that some of the larger models trained with pipeline parallelism can already train at peak efficiency with only hundreds of Mb/s bandwidth.

In theory, the square-cube principle also applies to intra-layer parallelism, but using this technique at 500 Mb/s would become practical only for layer sizes of more than $2^{16}$ units. Data-parallel training with sharding or offloading~\citep{zerooffload} does not scale as well, as its communication time scales with the size of \textit{model parameters} instead of activations. However, it may be possible to achieve similar scaling with gradient compression algorithms.

\subsection{SWARM Parallelism}\label{sect:method_swarm}

Traditional pipeline parallelism can be communication-efficient, but this alone is not enough for our setups. Since training devices can have different compute and network capabilities, a pipeline formed out of such devices would be bottlenecked by the single ``weakest link'', i.e., the participant with the smallest training throughput. As a result, the more powerful nodes along the pipeline would be underutilized due to either lack of inputs or slow subsequent stages. On top of that, if any node fails or leaves training prematurely, it will stall the entire training procedure.

\begin{figure*}[t]
    \centering
    \includegraphics[width=\linewidth]{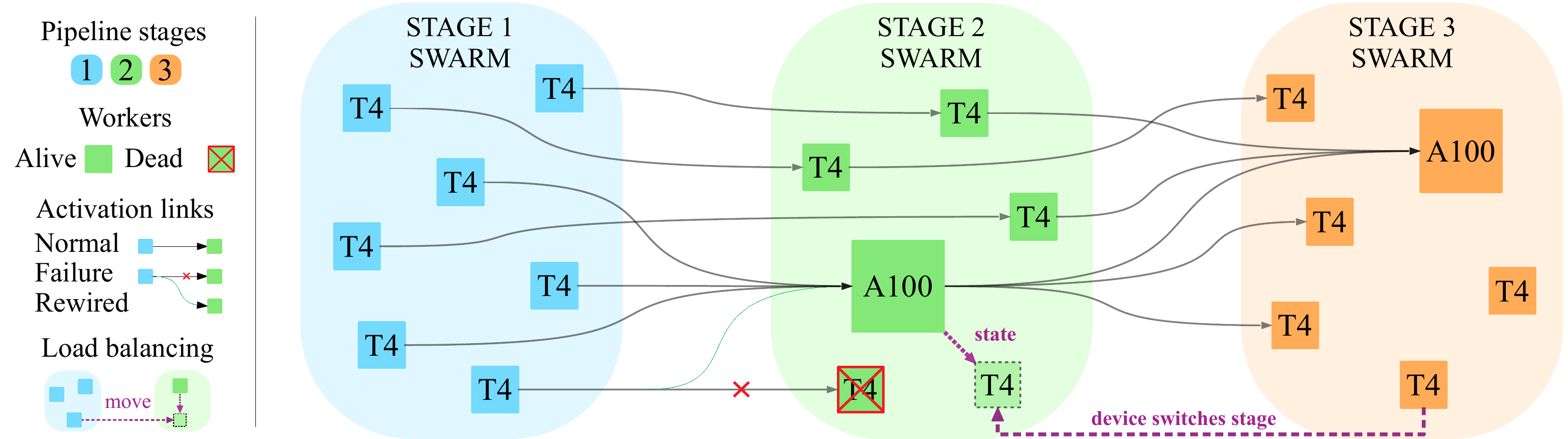}
    \caption{An overview of SWARM parallelism, illustrating both normal operation, device failures and adaptive rebalancing. One of the workers at stage 2 leaves; another peer from stage 3 takes its place by downloading the latest stage 2 parameters and statistics from peers.}
    \label{fig:swarm}
    \vspace{-12pt}
\end{figure*}

To overcome these two challenges, we replace the rigid pipeline structure with temporary ``pipelines'' that are built stochastically on the fly during each iteration. Each participant can send their outputs to any peer that serves the next pipeline stage. Thus, if one peer is faster than others, it can process inputs from multiple predecessors and distribute its outputs across several weaker peers to maximize utilization. Also, if any participant disconnects, its predecessors can reroute their requests to its neighbors. New peers can download up-to-date parameters and optimizer statistics from remaining workers at the chosen stage. This allows the training to proceed as long as there is at least one active participant per stage: we elaborate on the fault tolerance of SWARM parallelism in Appendix~\ref{appendix:faq}.

The resulting system consists of several consecutive swarms, as depicted in Figure~\ref{fig:swarm}. Peers within one swarm serve the same pipeline stage (i.e., the \textbf{same subset of layers} with \textbf{the same parameters}).
We assume that the model consists of similar ``blocks'' and thus partition it into evenly sized stages, leaving the study of better strategies~\citep{huang2019gpipe,pipedream} as future work.
During the \textit{forward} pass, peers receive inputs from predecessors (determined on each iteration) and send activations to peers in the next stage. For the \textit{backward} pass, peers receive gradients for outputs, compute gradients for layer inputs and accumulate gradients for parameters. Once enough gradients are accumulated, peers form groups, run All-Reduce to average gradients within their pipeline stages and perform the optimizer step.

SWARM parallelism can also use Delayed Parameter Updates~(DPU)~\citep{zerooffload} to further improve hardware utilization by performing the optimizer step in parallel with processing the next batch. While it is technically asynchronous, DPU was shown to achieve similar per-iteration convergence as fully synchronous training, both theoretically~\citep{stich2020error,arjevani2020tight} and empirically~\citep{zerooffload,dedloc}.

Each peer has queues for incoming and outgoing requests to maintain high GPU utilization under latency and to compensate for varying network speeds. Similarly to other pipeline implementations~\citep{huang2019gpipe,megatron2}, SWARM parallelism uses activation checkpointing~\citep{gradient_checkpointing_autograd, gradient_checkpointing_dl} to reduce the memory footprint. %

\vspace{-6pt}
\paragraph{Stochastic wiring.} To better utilize heterogeneous devices and recover from faults, we dynamically ``wire'' each input through each stage and pick devices in proportion to their training throughput. To achieve this, SWARM peers run ``trainer'' processes that route training data through the ``stages'' of SWARM, balancing the load between peers.

For each pipeline stage, trainers discover which peers currently serve this stage via a Distributed Hash Table (DHT,~\citealp{kademlia}). Trainers then assign a microbatch to one of those peers based on their performance. If that peer fails, it is temporarily banned and the microbatch is sent to another peer within the same stage. Note that trainers themselves do not use GPUs and have no trainable parameters, which makes it possible to run multiple trainers per peer. 

Each trainer assigns data independently using the Interleaved Weighted Round-Robin~\citep{iwrr,interleaved_round_robin} scheduler. Our specific implementation of IWRR uses a priority queue: each peer is associated with \textit{the total processing time over all previous requests}. A training minibatch is then routed to the node that has the smallest total processing time. Thus, for instance, if device A takes half as long to process a sample as device B, the routing algorithm will choose A twice as often as B. Finally, if a peer does not respond or fails to process the batch, trainer will ``ban'' this peer until it reannounces itself in the DHT, which is done every few minutes. For a more detailed description of stochastic wiring, please refer to Appendix~\ref{appendix:wiring_details}.

Curiously, different trainers can have different throughput estimates for the same device because of the network topology. For instance, if training nodes are split between two cloud regions, a given peer's trainer will have a higher throughput estimate for peers in the same data center. In other words, trainers automatically adjust to the network topology by routing more traffic to peers that are ``nearby''.

\vspace{-8pt}
\paragraph{Adaptive swarm rebalancing.} While stochastic wiring allows for automatic rebalancing within a stage, additional cross-stage rebalancing may be required to maximize throughput, especially when devices are very unreliable. As we described in Section~\ref{sect:related_cost_efficent_collaborative}, our workers can join and leave training at any time. If any single pipeline stage loses too many peers, the remaining ones will face an increased processing load, which will inevitably form a bottleneck. 

SWARM parallelism addresses this problem by allowing peers to dynamically switch between ``pipeline stages'' to maximize the training throughput. Every $T$ seconds, peers measure the utilization rate of each pipeline stage as the queue size.
Peers from the most underutilized pipeline stage will then switch to the most overutilized one (see Figure~\ref{fig:swarm} for an overview and Appendix~\ref{appendix:rebalancing_formal} for a formal description and complexity analysis), download the latest training state from their new neighbors and continue training. Similarly, if a new peer joins midway through training, it is assigned to the optimal pipeline stage by following the same protocol. As a side effect, if one pipeline stage requires more compute than others, SWARM will allocate more peers to that stage. In Section~\ref{sect:experiments_adaptive}, we evaluate our approach to dynamic rebalancing in realistic conditions.

\section{Experiments}

\subsection{Communication Efficiency at Scale}\label{sect:experiments_square_cube}

Before we can meaningfully evaluate SWARM parallelism, we must verify our theoretical observations on communication efficiency. Here we run several controlled experiments that measure the GPU utilization and network usage for different model sizes, using the Transformer architecture~\citep{transformer} that has been widely adopted in various fields~\citep{lin2021survey}. To decouple the performance impact from other factors, we run these experiments on homogeneous V100 GPU nodes that serve one pipeline stage over the network with varying latency and bandwidth. We use a batch size of 1 and sequences of 512 tokens; the complete configuration is deferred to Appendix~\ref{appendix:detailed_setup}.

First, we measure how the model size affects the computation to communication ratio at 500 Mb/s network bandwidth in both directions. We consider 4 model configurations: the base configuration from the BERT paper~\citep{bert}, ``xxlarge" (``large'' with $d_{model}{=}4096$),  which is used in several recent works~\citep{albert,ernie3,deberta}, and a GPT-3-scale model with $d_{model}{=}12288$~\citep{gpt3}. We also evaluate a modified Transformer architecture (``Ours'') as defined in Section~\ref{sect:experiments_large} with $d_{model}{=}4096$, 3 layers per pipeline stage and 8-bit quantized activations. As we demonstrate in Appendix~\ref{appendix:compression}, this compression strategy can significantly reduce network usage with little effect on convergence. In the first three configurations, the model consists of 12 Transformer layers placed on 12 servers with a single GPU; in the last one, there are 4 servers, each hosting 3 layers.
Appendix~\ref{appendix:detailed_setup} contains FLOP and parameter counts of each configuration.

\begin{figure}[b]
\vspace{-14pt}
    \centering
    \includegraphics[width=1\linewidth]{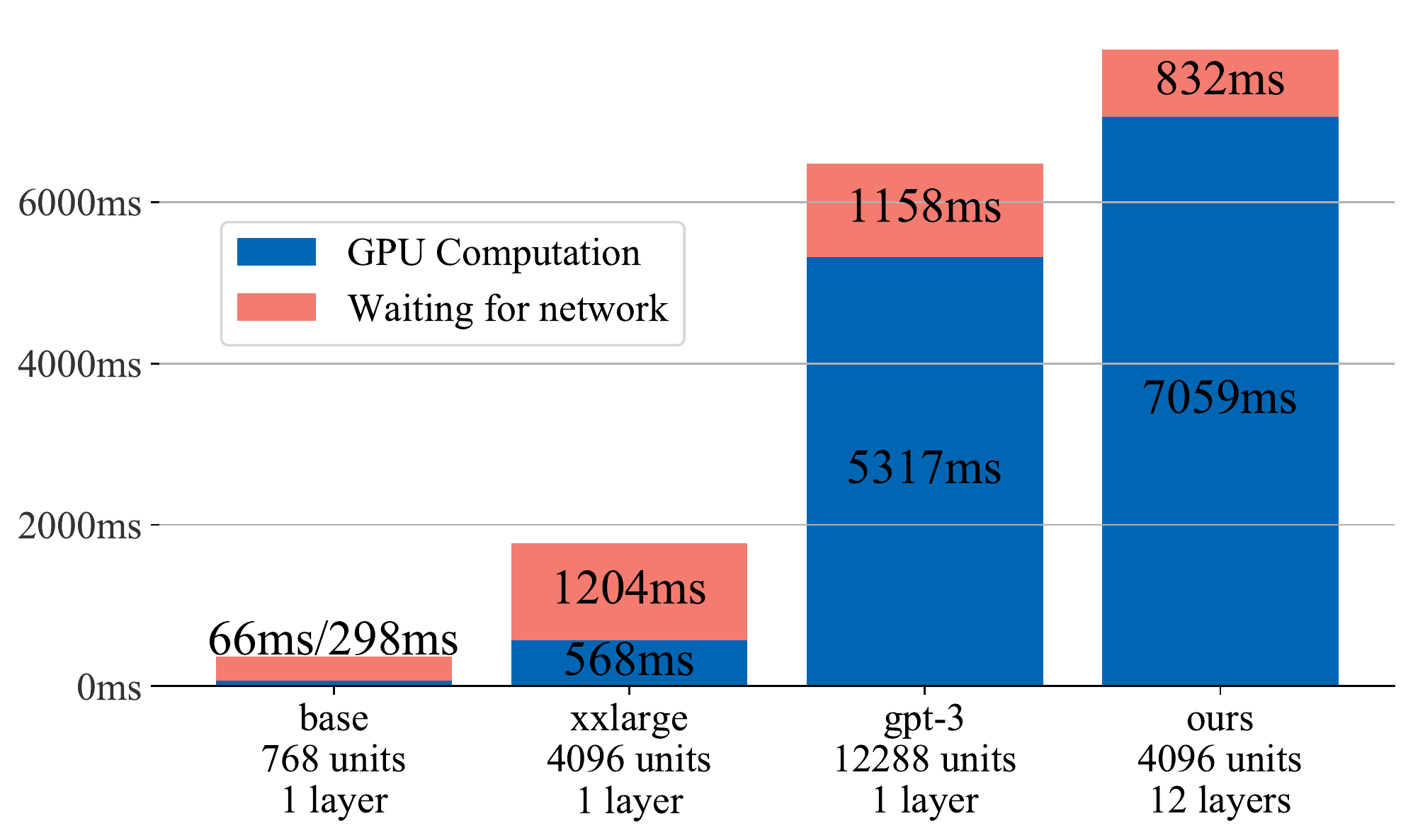}
    \vspace{-12pt}
    \captionof{figure}{Pipeline computation and idle time per batch at 500 Mb/s bandwidth.}
    \label{fig:throughput_exps}
\end{figure}%
\begin{table}
    \centering
    \captionof{table}{Relative device utilization at 500 Mb/s bandwidth and varying network latency.}
    \label{tab:latency}
    \small
    \setlength{\tabcolsep}{8pt}
    \begin{tabular}[b]{@{}lcccc@{}}
    \toprule
    \multirow{2}{*}{\thead{Latency\\(RTT)}} & 
    \multicolumn{4}{c}{
    \thead{
    Relative GPU utilization\\ (100\% - idle time)
    }
    }
    
    \\
\cmidrule{2-5}                     & base & xxlarge & GPT-3 & Ours \\ \midrule
    None &   18.0\%     &  32.1\%         &  82.1\%  &  89.5\%      \\
    10ms &   11.8\%      &   28.9\%    &   79.3\%   &  87.2\%    \\
    50ms &    4.88\%      &   20.1\%    &   70.3\% &  79.5\%    \\
    100ms &    2.78\%      &    14.9\%    &  60.2\%     &   71.5\% \\
    200ms &   1.53\%     &  10.1\%    &  48.5\%   &     59.2\%    \\
    \bottomrule
    \end{tabular}
    \vspace{-6pt}
\end{table}

As depicted in Figure~\ref{fig:squarecube} (right) and Figure~\ref{fig:throughput_exps}, larger models achieve better GPU utilization rate in the same network conditions, since their communication load grows slower than computation. More importantly, even at 500 Mb/s, the resulting GPU idle time can be pushed into the 10--20\% range, either naturally for GPT-3-sized models or through activation compression for smaller models. In addition, large models maintain most of their training efficiency at the 100ms latency~(Table~\ref{tab:latency}), which is roughly equivalent to training on different continents~\citep{verizon_latency}.

\vspace{-4pt}
\subsection{Detailed Performance Comparison}\label{appendix:training_throughput}

Here we investigate how SWARM parallelism compares to existing systems for training large models: \textbf{GPipe}~\citep{huang2019gpipe} and \textbf{ZeRO-Offload}~\citep{zerooffload}.
The purpose of this section is to compare the training throughput in ``ideal'' conditions (with homogeneous reliable devices and balanced layers), as deviating from these conditions makes it \textit{infeasible} to train with baseline systems.
Still, even in such conditions the performance of different systems can vary across model architectures, and hence we want to identify the cases in which using SWARM is preferable to other approaches.
We benchmark individual SWARM components in preemptible setups in Section~\ref{sect:experiments_adaptive} and Appendix~\ref{appendix:scaling}.

We evaluate training performance for sequences of 4 Transformer layers of identical size distributed over 16 workers. Similarly to Section~\ref{sect:experiments_square_cube}, we use three layer configurations: ``xxlarge''~($d_{model} {=} 4096$, $d_{\text{FFN}} {=} 16384$, 32 heads), ``GPT-3''~($d_{model} {=} 12288$, $d_{\text{FFN}} {=} 49152$, 96 heads), and ``Ours''~($d_{model} {=} 4096$, $d_{\text{FFN}} {=} 16384$, 32 heads, 16 shared layers per block, last stage holds only the vocabulary projection layer). The microbatch size is 4 for ``xxlarge'' and 1 for ``GPT-3'' and ``Ours'', and the sequence length is 512.

To provide a more detailed view of the training performance, we measure two separate performance statistics: the training throughput and the All-Reduce time. 
The training throughput measures the rate at which the system can process training sequences, i.e., run forward and backward passes. 
More specifically, we measure the time required to process 6250 sequences of 512 tokens, which corresponds to the largest batch size used in~\citet{gpt3}.
In turn, the All-Reduce time is the time each system spends to aggregate accumulated gradients across devices. 
Intuitively, training with small batch sizes is more sensitive to the All-Reduce time (since the algorithm needs to run All-Reduce more frequently) and vice versa.

\textbf{Hardware setup:} Each worker uses a V100-PCIe GPU with 16 CPU threads (E5 v5-2660v4) and 128 GB RAM. The only exception is for ZeRO-Offload with ``GPT-3'' layers, where we had to double the RAM size because the system required 190 gigabytes at peak. Similarly to Section~\ref{sect:experiments_square_cube}, each worker can communicate at a 500 Mb/s bandwidth for both upload and download for a total of 1 Gb/s.
In terms of network latency, we consider two setups: with \textbf{no latency}, where workers communicate normally within the same rack, and with \textbf{latency}, where we introduce additional $100\pm50$ms latency directly in the kernel\footnote{More specifically, \texttt{tc qdisc add dev <...> root netem delay 100ms 50ms}}.

\textbf{GPipe configuration:} We use a popular PyTorch-based implementation of GPipe\footnote{The source code is available at \url{https://github.com/kakaobrain/torchgpipe}}. The model is partitioned into 4 stages repeated over 4 model-parallel groups. To fit into the GPU memory for the ``GPT-3'' configuration, we offload the optimizer into RAM using ZeRO-Offload. Before averaging, we use PyTorch's built-in All-Reduce to aggregate gradients.
We evaluate both the standard GPipe schedule and the 1F1B schedule~\citep{pipedream}.

\textbf{ZeRO-Offload configuration:} Each worker runs the entire model individually, then exchanges gradients with peers. For ``xxlarge'', we use the official implementation from~\cite{zerooffload}. However, for ``GPT-3'', we found that optimizer offloading still does not allow us to fit 4 layers into the GPU. For this reason, we also offload the model parameters using the \texttt{offload\_param} option.

\begin{table}
\centering
\small
\setlength{\tabcolsep}{4pt}
\captionof{table}{Training performance for different model sizes.}
\label{tab:throughput_gpt}
\begin{tabular}[b]{lcccc}
\toprule
\multirow{2}[2]{*}{System} &
  \multicolumn{2}{c}{Throughput, min/batch} &
  \multicolumn{2}{c}{All-Reduce time, min} \\ \cmidrule(lr){2-3}\cmidrule(lr){4-5} 
                 & No latency & Latency & No latency & Latency \\
 \midrule \multicolumn{5}{c}{``GPT-3'' (4 layers) }\\
 \midrule
SWARM            &  168.3 &\textbf{186.7}  &  7.4 & \textbf{7.6}   \\
GPipe            &  164.5 & 218.4 &  \multirow{2}{*}{\textbf{6.7}}    & \multirow{2}{*}{7.8}   \\
1F1B & \textbf{163.3} & 216.1 & & \\
Offload          &  272.7 & 272.7          &  25.5 & 27.3 \\
\midrule \multicolumn{5}{c}{``xxlarge'' (4 layers) }\\
\midrule
SWARM            &  44.2 & 48.2                  &  0.8  & \textbf{0.9}   \\
GPipe            &  40.1 & 108.8                  &  \multirow{2}{*}{\textbf{0.7}}  & \multirow{2}{*}{1.1}   \\
1F1B & 40.8 & 105.5 & & \\
Offload          &  \textbf{33.8} & \textbf{33.8}  &  2.8 & 4.2   \\
\midrule \multicolumn{5}{c}{Full ``Ours'' model (48 shared layers + embeddings) }\\
\midrule
SWARM            &  432.2 & 452.9                  &  0.8  &\bf 1.0   \\
GPipe            &  420.0 & 602.1                   &  \multirow{2}{*}{\bf 0.7}  & \multirow{2}{*}{1.1}   \\
1F1B             &  408.5 & 569.2 & & \\
Offload          &  \bf 372.0 &\bf 372.0  &  3.2 & 4.8   \\
\bottomrule
\end{tabular}
\vspace{-8pt}
\end{table}%

\begin{figure}[b]
\vspace{-16pt}
\centering
\includegraphics[ width=0.65\linewidth]{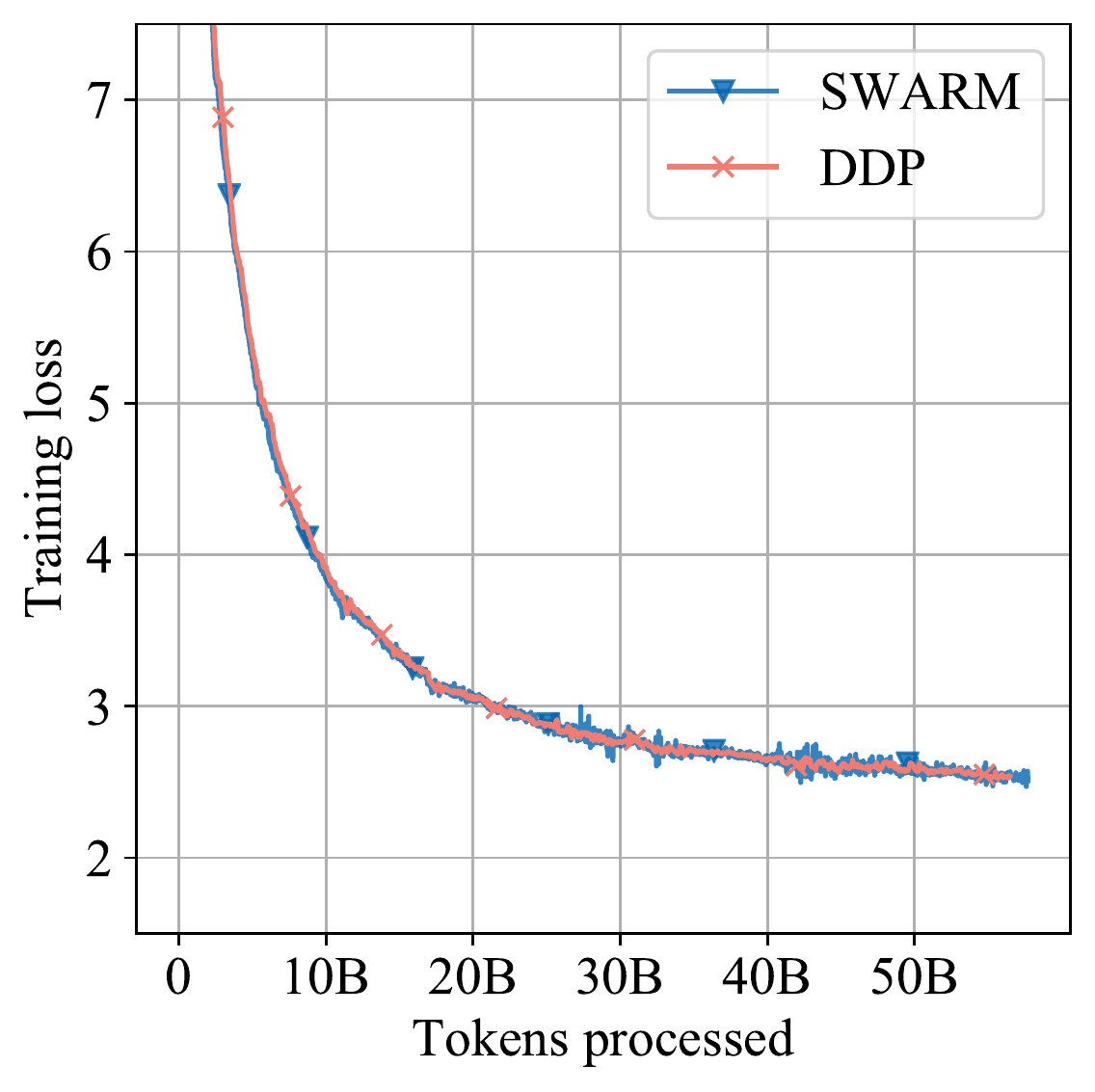}
\vspace{-6pt}
\captionof{figure}{Training convergence comparison.}
\label{fig:convergence}
\end{figure}

In turn, when training smaller models, ZeRO-Offload outperforms both SWARM and GPipe. This result aligns with our earlier observations in Figure~\ref{fig:squarecube}, where the same model spent most of the time waiting for the communication between pipeline stages.%

We also observe that ZeRO-Offload takes longer to aggregate gradients, likely because each peer must aggregate the entire model, whereas in SWARM and GPipe, peers aggregate a single pipeline stage. The variation between All-Reduce time in GPipe and SWARM is due to implementation differences. Overall, SWARM is competitive to HPC baselines even in an idealized homogeneous environment.

\subsection{Large-Scale Distributed Training}
\label{sect:experiments_large}

To verify the efficiency of SWARM parallelism in a practical scenario, we conduct a series of large-scale distributed experiments using preemptible (unreliable) cloud T4 and A100 GPUs over a public cloud network.

We train a Transformer language model with the architecture similar to prior work~\citep{gpt3,gptj,gptneo} and 1.01 billion parameters in total. Our model consists of 3 stages, each containing a single Transformer decoder block with $d_{model}=4096$ and 16 layers per pipeline stage. All workers within a stage serve the same group of layers, and all layers within each group use the same set of parameters, similarly to ALBERT~\citep{albert}. On top of this, the first stage also contains the embedding layer, and the last stage includes the language modeling head. Because of layer sharing, this model is equivalent to a 13B model from~\citet{gpt3} in terms of compute costs. 

We use 8-bit compression~\citep{adam8bit} for activations and gradients to reduce the communication intensity. Additional training setup details are covered in Appendix~\ref{appendix:detailed_large}.
SWARM nodes run rebalancing every $T=300$ seconds, and trainers measure peer performance using a moving average with $\alpha=0.1$. However, as we show in Section~\ref{sect:experiments_adaptive}, the throughput of SWARM is not very sensitive to the choice of these hyperparameters.

First, to verify that model parallelism with asynchronous updates does not have significant convergence issues, we train the model on the Pile~\citep{gao2020pile} dataset with 400 preemptible T4 instances, each hosting one accelerator. As a baseline, we use regular data-parallel training with offloading on 128 A100 GPUs.
We run both experiments for approximately 4 weeks and compare the learning curves.

Figure~\ref{fig:convergence} shows the results of this experiment: it can be seen that the training dynamics of two approaches are indeed similar, which demonstrates the viability of SWARM parallelism for heterogeneous and poorly-connected devices.

In the next experiment, we aim to measure the pipeline throughput in different hardware conditions and to compare it with an estimate of best-case pipeline performance.
We consider several setups: first, we use the same 400 preemptible T4 nodes; in another setup, we use 7 instances with 8 A100 GPU each; finally, we combine these fleets to create a heterogeneous setup. We examine the performance of the pipeline both with weight sharing and with standard, more common, Transformer blocks.

\begin{table}
\centering
\captionof{table}{Pipeline throughput, layer sharing.}
\label{tab:throughput}
\small
\begin{tabular}{@{}lcccc@{}}
\toprule
\multirow{2}{*}{\begin{tabular}[c]{@{}l@{}}Hardware\\ setup\end{tabular}} &
  \multicolumn{2}{c}{\begin{tabular}[c]{@{}c@{}}Throughput,\\ samples/s\end{tabular}} &
  \multicolumn{2}{c}{\begin{tabular}[c]{@{}c@{}}Optimal\\ bandwidth, Mb/s\end{tabular}} \\ \cmidrule(lr){2-3}\cmidrule(lr){4-5} 
                 & Actual & Best-case & Upload & Download \\ \midrule
T4           &  17.6      &   19.2        &   317.8     &     397.9     \\
A100          & 16.9       &   25.5        &   436.1     &     545.1     \\
T4 \& A100 &   27.3     &       ---    &   ---     &      ---    \\ \bottomrule
\end{tabular}
\end{table}
\begin{table}
\centering
\captionof{table}{Pipeline throughput, default Transformer.}
\label{tab:throughput_standard}
\small
\begin{tabular}{@{}lcc@{}}
\toprule
\multirow{2}{*}{\begin{tabular}[c]{@{}l@{}}Hardware\\ setup\end{tabular}} &
  \multicolumn{2}{c}{\begin{tabular}[c]{@{}c@{}}Throughput,\\ samples/s\end{tabular}} \\ \cmidrule(lr){2-3}
                 & Actual & Best-case \\ \midrule
T4           &  8.8      &   19.3        \\
A100          & 8.0       &   25.1        \\
T4 \& A100 &   13.4     &       ---    \\ \bottomrule
\end{tabular}
\end{table}

We measure the number of randomly generated samples processed by the pipeline both in our infrastructure and the ideal case that ignores all network-related operations (i.e., has infinite bandwidth and zero latency). The ideal case is emulated by executing a single pipeline stage 3 times locally on a single server and multiplying the single-node estimates by the number of nodes.

As demonstrated in the left two columns of Table~\ref{tab:throughput} and Table~\ref{tab:throughput_standard}, asynchronous training of compute-intensive models with 8-bit compressed activations regardless of the architecture specifics allows us to achieve high performance without a dedicated networking solution. Furthermore, the load balancing algorithm of SWARM allows us to dynamically and efficiently utilize different hardware without being bottlenecked by slower devices.

Next, we use the same load testing scenario to estimate the bandwidth required to fully utilize each device type in the above infrastructure. For this, we measure the average incoming and outgoing bandwidth on the nodes that serve the intermediate stage of the pipeline. We summarize our findings in the right two columns of Table~\ref{tab:throughput}: it turns out that with layer sharing and 8-bit compression, medium-performance GPUs (such as T4) can be saturated even with moderate network speeds. Based on our main experiment, the optimal total bandwidth is roughly 100Mb/s higher than the values reported in Table 3 due to gradient averaging, loading state from peers, maintaining the DHT and streaming the training data.
Although training over the Internet with more efficient hardware might indeed underutilize the accelerator, this issue can be offset by advanced compression strategies such as compression-aware architectures or layer sharing, as shown in Table~\ref{tab:throughput}.

\begin{figure}[t]
    \centering
    \includegraphics[width=\linewidth]{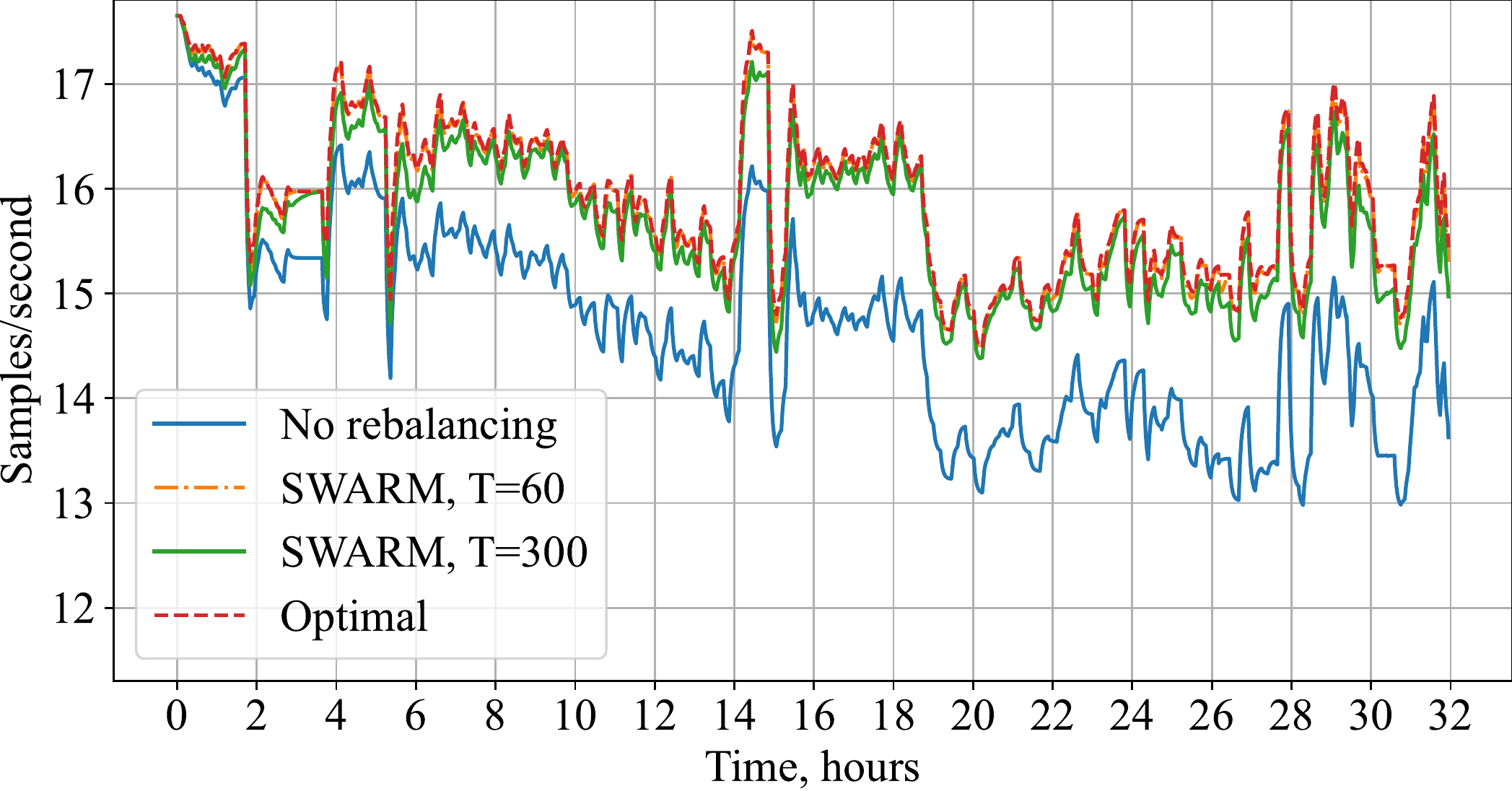}
    \captionof{figure}{Throughput of rebalancing methods over time.}
    \label{fig:rebalancing}
\end{figure}

\subsection{Adaptive Rebalancing Evaluation}

\begin{figure*}[h!]
\begin{subfigure}{0.5\textwidth}
    \centering
    \includegraphics[width=0.97\linewidth]{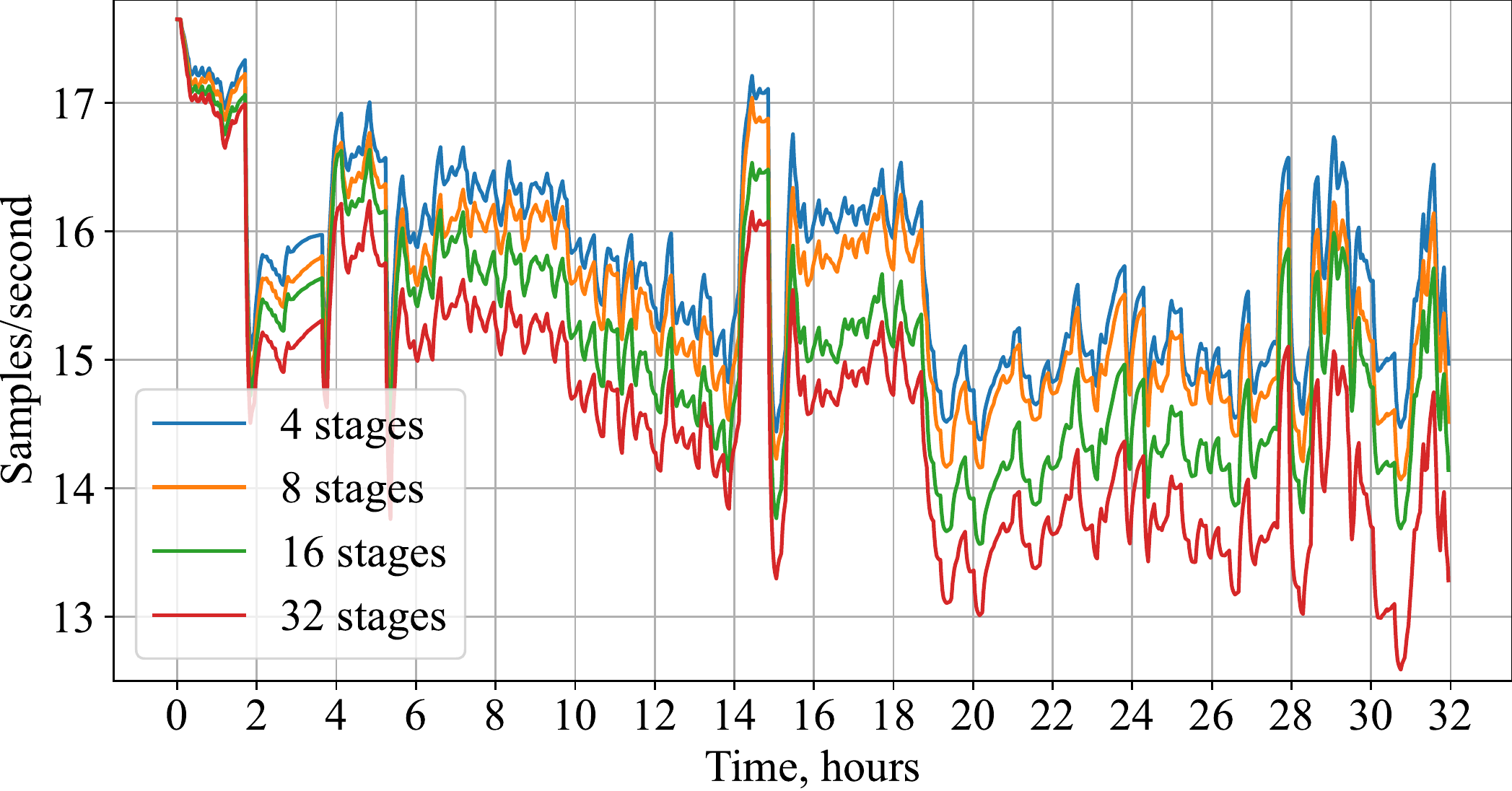}
    \caption{Adaptive rebalancing of SWARM parallelism.}
    \label{fig:rebalancing_stages}
\end{subfigure}%
\begin{subfigure}{0.5\textwidth}
    \centering
    \includegraphics[width=0.97\linewidth]{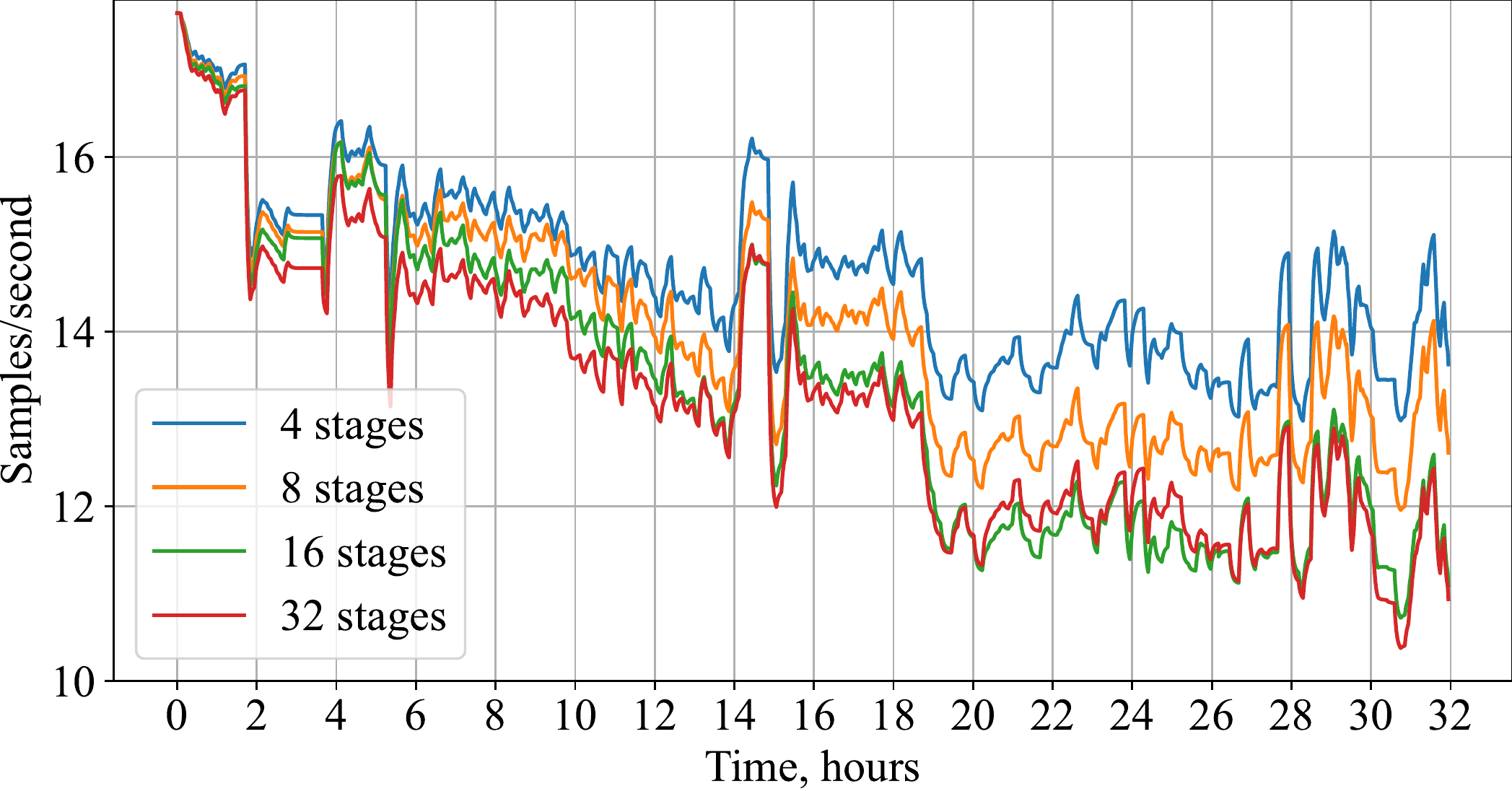}
    \caption{No rebalancing.}
    \label{fig:rebalancing_stages_baseline}
\end{subfigure}
\caption{Scaling of pipeline-parallel strategies with respect to the number of stages.}
\label{fig:rebalancing_stages_all}
\end{figure*}

\label{sect:experiments_adaptive}
In this experiment, we evaluate the efficiency of adaptive peer rebalancing between stages proposed in Section~\ref{sect:method_swarm}. 
We use statistics of the number of active T4 nodes from the 32-hour segment of the experiment described in Section~\ref{sect:experiments_large}. 
We use this data to simulate training dynamics by viewing it as sequence of events, each consisting of a timestamp and a change in the number of peers (which can be positive or negative). 
When a worker is removed from the pipeline, we randomly choose the stage it was removed from: that is, removing $N$ peers corresponds to $N$ samples from the uniform distribution over four pipeline stages. 
We run 10 simulations with different random seeds and average the resulting trajectories.
We compare our strategy with two different values of $T$ to the baseline that has no rebalancing.

The results of this evaluation are available in \autoref{fig:rebalancing}; for reference, we also provide the performance of a theoretically optimal rebalancing strategy that maintains the highest possible throughput at every moment. It can be seen that even with the rebalancing period $T=300$, our approach significantly improves the overall throughput of the pipeline. When the number of peers is relatively stable, the rebalanced pipeline also approaches the optimal one in terms of throughput, which shows the efficiency of rebalancing even when moving only one node at a time.

In addition, we observed that for some brief periods, the performance of the unbalanced pipeline exceeded the throughput of the balanced one due to random choice of disconnecting peers (dropping more from the ``overrepresented'' stages affects the imbalanced pipeline less). However, this held true only for $\approx 4.5\%$ of the experiment and was quickly mitigated by adaptive rebalancing.

As expected, decreasing $T$ from 300 to 60 seconds improves both the overall throughput and the speed of convergence to optimal pipeline performance. However, the effect is not as drastic compared to the increase in DHT data transfer volume. This is also demonstrated by \autoref{tab:rebalancing_speedup}, which shows the relative throughput of the three configurations compared to the optimal one. Furthermore, the table displays that while initially there is little difference between rebalancing choices, it becomes more pronounced later on as the imbalanced version ``drifts further'' from the optimal state.

\begin{table}[b]
\centering
\captionof{table}{Relative throughput comparison of pipeline rebalancing methods.}
\small
\label{tab:rebalancing_speedup}
\begin{tabular}[b]{@{}lccc@{}}
\toprule
\multirow{2}[2]{*}{\thead{Rebalancing}} & \multicolumn{3}{c}{\% of optimal} \\ \cmidrule(l){2-4} 
                        & Overall   & First 1 hour   & Last 1 hour  \\ \midrule
None                & 82.7      & 99.0       & 45.4     \\
$T=300$    & 95.8      & 99.4       & 88.9     \\
$T=60$     & 97.6      & 99.8       & 91.7     \\ \bottomrule
\end{tabular}
\end{table}

Finally, we analyze the scaling properties of rebalancing with respect to the number of stages. To do this, we conduct experiments in the same setup as above ($T=300$) while changing the number of pipeline stages from 4 to $\{4,\ 8,\ 16,\ 32\}$. To ensure the consistency of throughput across all experiments, we increase the starting number of peers accordingly while keeping the preemption rate constant. As a baseline, we also evaluate the throughput of the pipeline that has no rebalancing.

Figure~\ref{fig:rebalancing_stages_all} shows the outcome of this experiment. As displayed in the plots, both strategies drop in performance with the increase in the stage count: while all stages should drop in performance equally in expectation, in practice, the variances are too large while the number of peers is relatively too small for the asymptotic properties to take place. This effect results in more outliers (large drops in the number of peers) in the preemption distribution for more stages. Still, rebalancing allows to partially mitigate the issue: while we observe a more consistent downward trend for the baseline strategy, the rebalanced pipeline regains its performance over time and achieves a higher overall throughput.

\section{Conclusion}

In this work, we evaluate the feasibility of high-throughput training of billion-scale neural networks on unreliable peers with low network bandwidth. 
We find that training in this setup can be possible with very large models and pipeline parallelism.
To this end, we propose SWARM parallelism to overcome the challenges of pipeline parallelism for preemptible devices with heterogeneous network bandwidths and computational throughputs. 
We show that our method is highly effective at rebalancing peers and maximizing the aggregate training throughput even in presence of unstable nodes.
We also show that training \textbf{large models} with \textbf{SWARM parallelism} and \textbf{compression}-aware architectures enables high utilization of cheap preemptible instances with slow interconnect. 
As such, our work makes training of large models accessible to researchers that do not have access to dedicated compute infrastructure.

\pagebreak

\bibliographystyle{icml2023}
\bibliography{bibliography}

\clearpage
\appendix
\section*{Supplementary Material}

This part of the paper is organized as follows:
\begin{itemize}
    \item \cref{appendix:faq} overviews several common questions about the details of our study and addresses the limitations of SWARM parallelism;
    \item In \cref{appendix:related}, we list further related works on topics relevant to the problem setting we study;
    \item In \cref{appendix:wiring_details} and \cref{appendix:rebalancing_formal}, we give a more formal description and outline the details of stochastic wiring and adaptive rebalancing, accordingly;
    \item In \cref{appendix:equivalence}, we outline the relation between training with SWARM and using methods for offloading.
    \item \cref{appendix:detailed_setup} and  \cref{appendix:detailed_large} contain additional details of our experimental setup, whereas \cref{appendix:scaling} reports further experiments on specific aspects and components of SWARM parallelism;
    \item Lastly, we investigate compression-aware architectures in \cref{appendix:compression} and evaluate their impact in a practical setting in \cref{appendix:time_to_solution}.
\end{itemize}

\section{Answers to Common Questions}
\label{appendix:faq}

\paragraph{Why not just use data parallelism with offloading?}

Regular data parallelism requires all-reduce steps where peers exchange gradients, which can be prohibitively expensive for large models. For example, a 1 billion parameter model with 16-bit gradients requires 2 GB of data to be synchronized between all $n$ devices. We need at least $n$ messages to perform this synchronization. If we have 100 devices with bidirectional communication, each client would need to send 2 GB of data to finish the synchronization. Thus, with slow interconnects, such synchronizations are not practical.

\paragraph{Why not just use fully sharded data parallelism with elasticity?}

Sharded data parallelism requires all-to-all communication of parameter buffers at each layer. Each of these communications can be done in parallel and has a size of parameter count divided by $n$; in total, $n$ messages are required. Thus, for 1B parameters in 16-bit precision, a total of 2 GB need to be synchronized for both the forward and backward pass. For low-bandwidth devices with 100 Mb/s speed, this would entail an overhead of 5.5 minutes per forward/backward pass, which is difficult to overlap with computation. This is exacerbated further, because all-to-all communication latency is determined by the slowest peer. Thus, sharded data parallelism can be particularly inefficient for setups where peers have different network bandwidths.

\paragraph{Should I use SWARM in a supercomputer?}
By default, SWARM is worse than traditional parallelism due to its extra complexity (see experiments in Section~\ref{appendix:training_throughput}). However, SWARM can be useful in case of supercomputers that have heterogeneous devices.

\paragraph{ZeRO-Offload allows one to train 13B parameters on a single V100, so why do I need SWARM?}

Using ZeRO-Offload can slow down training due to the slow data transfer between external memory and the accelerator. Training with SWARM can {\it accelerate} training while also allowing to train larger models; see Appendix~\ref{appendix:equivalence} for a detailed comparison.

\paragraph{Is it worth using preemptible instances and SWARM from an economic standpoint?}
Due to a significantly smaller cost per hour, one can leverage a larger amount of computation when using spot instances compared to on-demand cloud VMs or dedicated HPC setups. See \autoref{appendix:time_to_solution} and \autoref{tab:cost} for a comparison of both hourly and total costs for an example large-scale pretraining task.

\paragraph{When should I avoid using SWARM?}

SWARM is efficient at training compute-intensive models with more than 1B parameters. For smaller models, a sharded data-parallel approach can be more optimal. For homogeneous HPC environments, standard sharded data-parallel or pipeline-parallel training will be more efficient than SWARM, because the rebalancing is not required. For HPC environments that are so extensive that the failure of a node is likely, the practicality of SWARM depends on how many nodes are expected to fail. Elastic sharded data parallelism is better than SWARM if the number of expected failures is relatively low.

\paragraph{Can I use SWARM without layer sharing or quantization?}
Yes, SWARM can still be effective in these scenarios. Our bandwidth experiments in the main part of the work give an estimate of its network overhead. By using no quantization, which means using regular 16-bit activations, the network overhead increases approximately by a factor of two. Without layer sharing, the overhead within each pipeline stage to synchronize the gradients is increased by the number of layers not being shared. As such, a rough estimate of the efficiency of SWARM in these scenarios can be estimated by taking our model size and network bandwidth requirements data and multiplying it by the relevant factor.

\paragraph{Do the compression-aware architecture modifications apply only to Transformers?} 
Bottleneck and maxout compression are general compression techniques that can be applied to any layer in any architecture. However, their effectiveness may vary depending on where in the model they are applied and what kind of model these are applied to (for example, CNNs vs. RNNs vs. Transformers).

\paragraph{How many pipeline stages can SWARM have?}
While its design allows for any number of stages, using long pipelines can result in a reduced training throughput. Similarly to regular pipeline parallelism, SWARM suffers from the pipeline ``bubble'' problem~\citep{huang2019gpipe}: at the beginning of the initial batch processing, peers near the end of the pipeline will be waiting for inputs. Likewise, early layers will be idle after processing the final microbatch.
In theory, this can be mitigated with asynchronous updates~\citep{pipedream,pipemare}, but we did not investigate them in this work due to potential convergence issues.

\paragraph{How much failure can SWARM handle?}
As long as there is at least one operational peer at every pipeline stage and at least one trainer, SWARM can work without any issues. 
The key factors defining the training run state at a given SGD step are the model parameters, the optimizer statistics, the data loader state, and the step number (required for proper scheduling). The up-to-date parameters and optimizer statistics, as well as the step number, are naturally located on all active nodes of a given stage, since they are required for training. Thus, when a peer joins the network, it can download the checkpoint corresponding to the current training state from other peers.

As we mention in Section~\ref{sect:method_swarm}, peer failures do not affect forward and backward passes as long as there is at least one peer at the required stage: because of rewiring, it is possible to resend activations or gradients to another worker that has identical model weights by construction. Similarly, the data loader state can be recomputed from the last known SGD step. However, we do not track the order of examples sampled within the same batch; because of the i.i.d. assumption in the large-scale training setup, the distribution of gradients is expected to be the same. Hence, if the peer leaves from the pipeline stage, other workers can compute gradients and replace those accumulated by the disconnected peer, so that the number of examples for an SGD step stays the same.

\paragraph{Some configurations in Section~\ref{sect:experiments_square_cube} measure less than $\bf 20\%$ GPU idle time, while many HPC systems only achieve $\bf \approx80\%$ GPU utilization. Does this mean that SWARM is $\bf 30\%$ faster?} 
No, because these are different measurement types. \citet{megatron2} measures GPU utilization as a fraction of theoretical peak FLOP/s of their GPUs. In contrast, we only measure what fraction of time the GPU is running the model, regardless of efficiency. Since any realistic deep learning workload cannot achieve $100\%$ peak FLOP/s, $20\%$ GPU idle time for SWARM means that it can reach $\approx0.8$x the training throughput compared to training with an infinitely fast network. As a rule of thumb, one can say that SWARM will run at a $20\%$ slower speed than systems described by~\citet{megatron2} using the infrastructure that is several times cheaper.%

\section{Additional Related Work}\label{appendix:related}

\vspace{-2pt}
\paragraph{Dynamic parameter loading.} Several recent studies propose alternative execution algorithms that allow training large models with data parallelism. Since neural networks typically use a small fraction of weights at any given moment, the remaining ``inactive'' parameters can be sharded~\citep{zero} or offloaded to external memory~\citep{l2l,zerooffload,zero_ssd}. In sharded data parallelism~\cite{zero}, inactive tensors are distributed across all $n$ devices such that each device stores $\frac{1}{n}$th of all parameters. For active layers, the shards are gathered such that each device holds the entire tensor just-in-time for computation. After the computation, the parameters' memory is freed so that only the sharded memory remains ($\frac{1}{n}$th per device). This makes it very memory efficient to store model and optimizer states for inactive layers if many devices are available. Similarly to tensor parallelism, these algorithms can support arbitrary models without the need for layer partitioning and can, in principle, run a large model on a single GPU, which is useful for finetuning and inference.

\vspace{-6pt}
\paragraph{Architecture-specific methods.} Finally, some distributed training algorithms take advantage of specific layers, such as locally connected layers~\citep{dean12,coates13}, Mixture-of-Experts~\citep{moe_first,shazeer2017outrageously,Lepikhin2020GShardSG}, Switch layers~\citep{fedus2021switch} or Product Key Memory~\citep{pkm}. These layers contain many near-independent parts that can be assigned to different devices. They can easily scale to an extremely large number of parameters with a relatively small increase in compute~\citep{shazeer2017outrageously}. However, they are also less parameter-efficient~\citep{fedus2021switch} and may not apply to all architectures.

\vspace{-6pt}
\paragraph{Optimal scheduling for distributed training.}
When the configuration of each peer is known, it is possible to significantly optimize the pipeline scheduling by going beyond the greedy approach with global optimization techniques~\citep{alpa,piper}, even with heterogeneous hardware~\citep{yuan2022decentralized}.
However, we consider a setup in which this is not possible: preemptible and volunteer peers can join at any point of the experiment, and dynamically rescheduling and orchestrating them in a centralized manner is technically difficult because of the communication and reliability constraints.

\vspace{-6pt}
\paragraph{Elastic training.} To train with a dynamic number of workers, deep learning practitioners have developed elastic training algorithms~\citep{pytorch_elastic,elastic_horovod}. If a worker leaves or fails during training, these algorithms rebalance the load between the remaining nodes and continue the training procedure~\citep{proteus,moshpit}. If new workers join during training, they get the latest model parameters from their peers and train alongside them.

\vspace{-10pt}
\paragraph{Asynchronous training.} Another important problem is distributed training on devices with uneven performance. One way to solve this problem is to use asynchronous training, where nodes compute gradients at their own pace and aggregate them using a parameter server~\citep{recht2011hogwild,volunteer_dl_async} or a decentralized network~\citep{dp_sgd}. This idea allows full utilization of each device, but may reduce the convergence rate due to ``stale'' gradients~\citep{recht2011hogwild,aji2019making}. Several studies~\citep{wagma,moshpit,zerooffload,dedloc} propose hybrid techniques that remove some synchronization points while maintaining the per-iteration convergence.

\section{Stochastic Wiring Details}\label{appendix:wiring_details}

Our approach uses \textit{stochastic wiring}, a specialized routing algorithm designed around heterogeneous unreliable devices and high network latency. The core idea of stochastic wiring is to route each training microbatch through random devices from each pipeline stage, such that the workload of each device is proportional to its performance.
The performance of the peer is measured as an exponentially weighted average of its response time, and all peers serving a specific stage are stored in a priority queue. 
We formally describe the components of stochastic wiring in Algorithm~\ref{alg:wiring}.

From a system design perspective, each worker runs a separate \textit{trainer} process that forms microbatches and routes them through pipeline stages (forward and backward pass). As we describe earlier in Section~\ref{sect:method_swarm}, trainers run Interleaved Weighted Round Robin~\citep{iwrr,interleaved_round_robin} (IWRR) scheduling to dynamically assign microbatches to peers based on each peer's training throughput (``samples per second'') in a balanced way.

An important observation is that \textit{stochastic wiring allows SWARM to mitigate network latency}. Unlike existing pipeline algorithms~\citep{huang2019gpipe}, SWARM workers do not get blocked if their neighbors take too long to process a minibatch. Instead, each SWARM device maintains a queue of microbatches assigned by trainers. In case of a latency spike, workers keep processing previously queued microbatches, maintaining high device utilization.

\begin{figure}[t]
\vspace{-1em}
\begin{algorithm}[H]
  \captionof{algorithm}{Pseudocode of stochastic wiring}
  \label{alg:wiring}
\begin{algorithmic}[1]
  \INPUT the number of pipeline stages $N$, the set of active servers $S$, smoothing parameter $\gamma$, initial priority $\epsilon$

  \STATE \(\triangleright\) Initialization
  \STATE ema = dict()
  \STATE queues = list()
  \FOR{$\text{i} \in 1,\ldots, N$}
  \STATE queues.append(PriorityQueue())
  \ENDFOR
  \STATE \textbf{def} add\_server(server)\textbf{:}
      \STATE \hspace{12px} ema[server] = $\varepsilon$
      \STATE \hspace{12px} \textbf{for} $\text{i} \in \text{get\_blocks\_served\_by(server)}$\textbf{:}
        \STATE \hspace{24px} queues[i].update(server, priority=$\varepsilon$)
  \STATE \textbf{def} ban\_server(server) \textbf{:}
      \STATE \hspace{12px} \textbf{for} $\text{i} \in \text{get\_blocks\_served\_by(server)}$\textbf{:}
        \STATE \hspace{24px} queues[i].update(server, priority=$\infty$)
  \STATE \textbf{def} choose\_server(i)\textbf{:}
      \STATE \hspace{12px} server, priority = queues[i].top()
      \STATE \hspace{12px} new\_priority = priority + ema[server]
      \STATE \hspace{12px} \textbf{for} $\text{j} \in \text{get\_blocks\_served\_by(server)}$ \textbf{:}
          \STATE \hspace{24px} queues[j].update(server, priority=new\_priority)
      \STATE \hspace{12px} \textbf{return} server
  \STATE \(\triangleright\) Forward pass with stochastic wiring
  \STATE \textbf{def} forward(inputs):
      \STATE \hspace{12px} layer\_index = 0
      \STATE \hspace{12px} \textbf{while} $\text{layer\_index} < N$\textbf{:}
          \STATE \hspace{24px} server = choose\_server(layer\_index)
          \STATE \hspace{24px} t = get\_current\_time()
          \STATE \hspace{24px} \textbf{try:}
          \STATE \hspace{36px} inputs = server.forward(inputs)
          \STATE \hspace{36px} layer\_index = layer\_index + 1
          \STATE \hspace{36px} $\Delta t$ = get\_current\_time() - t
          \STATE \hspace{36px} ema[server] = $\gamma \cdot \Delta t + (1 - \gamma) \cdot$ ema[server]
          \STATE \hspace{24px} \textbf{catch} (ServerFault, Timeout):
          \STATE \hspace{36px} ban\_server(server)
      \STATE \hspace{12px} \textbf{return} inputs
\end{algorithmic}
\end{algorithm}
\vspace{-25pt}
\end{figure}

\section{Description and Complexity of Adaptive Rebalancing}
\label{appendix:rebalancing_formal}

Algorithm~\ref{alg:adaptive_rebalancing} contains the formal definition of the adaptive rebalancing procedure. As described previously, each worker of SWARM that hosts model layers continuously updates the information about its load in parallel with processing the incoming requests. Each $T$ seconds, the peers measure the total load for all stages of the pipeline, and the peer with the lowest queue size from the stage with the minimum load moves to the stage with the maximum load. In principle, the algorithm could be extended to support moving multiple peers simultaneously; however, as we have shown in Section~\ref{sect:experiments_adaptive}, even in the current form the algorithm bridges most of the gap between the optimally balanced pipeline and the system without any rebalancing.

The complexity of Algorithm~\ref{alg:adaptive_rebalancing} can be estimated as follows: for $M$ as the highest number of peers over all stages, we have $O(M)$ operations in Lines 9--11 and Lines 22--24, and all other operations take constant time for a single stage. These operations are nested in the loop over all stages, which means that the total complexity of the algorithm is $O(MS)$. For practical numbers of both peers (e.g., < 10,000) and stages (fewer than 100), this incurs a negligible overhead on performance, as all communication and computation is done in parallel with the actual forward and backward passes.

Also, notice that only one migrating peer needs to stop processing requests and download the weights and optimizer statistics of the pipeline stage it starts serving: this means that the overall network load of this procedure is relatively small, as all DHT requests handle scalar data and do not exceed the number of active peers for each worker.

In practice, the algorithm handles slight deviations in local time and network/DHT latencies by allowing the peers to wait for straggling nodes in Line 9 for a predefined timeout. If a node does not join the rebalancing procedure by reporting its load in time or joins the network too late, it is omitted from the current iteration.

\begin{algorithm}
  \caption{Adaptive rebalancing for SWARM parallelism}
  \label{alg:adaptive_rebalancing}
\begin{algorithmic}[1]
  \INPUT peer index $i$, current peer stage $s_{cur}$, total number of stages $S$, rebalancing period $T$
  \WHILE{active}
  \STATE Sleep for $T$ seconds
  \STATE Measure $q_i$ as the local request queue size
  \STATE Write $(i, q_i)$ as the key-subkey pair to DHT[$s_{cur}$]
  \STATE Initialize minimum and maximum load stages: $s_{min}=s_{max}:=-1$,
  \STATE $l_{min}:=\infty, l_{max}:=-\infty$
  \FOR{$s$ in $1,\ldots, S$}
  \STATE Initialize the load buffer $L = 0$
  \FOR{$(j,q_j)$ in DHT[$s$]}
  \STATE $L:=L+q_j$
  \ENDFOR 
  \IF{$L>L_{max}$}
  \STATE $s_{max}:=s,\ L_{max}:=L$
  \ENDIF
  \IF{$L<L_{min}$}
  \STATE $s_{min}:=s,\ L_{min}:=L$
  \ENDIF
  \ENDFOR
  \IF{$s_{cur}=s_{min}$}
  \STATE // Migrate to the maximum load stage
  \STATE Initialize the minimum load peer $i_{min}:=-1,q_{min}:=\infty$
  \FOR{$(j,q_j)$ in DHT[$s$]}
  \IF{$q_j<q_{min}$}
  \STATE $i_{min}:=j,\ q_{min}:=q_j$
  \ENDIF
  \ENDFOR
  \IF{$i_{min}=i$}
  \STATE // This peer should migrate
  \STATE $s_{cur}:=s_{max}$
  \STATE Download up-to-date parameters from peers in $s_{max}$
  \ENDIF
  \ENDIF
  \ENDWHILE
\end{algorithmic}
\end{algorithm}

\section{Relation between SWARM and ZeRO-Offload}\label{appendix:equivalence}
\vspace{2pt}

In this section, we argue that depending on the use of DPU, SWARM-parallel training is equivalent to either fully synchronous training or the semi-synchronous training proposed in ZeRO-Offload~\citep{zerooffload}.
That is, SWARM produces exactly the same stepwise updates as conventional distributed training algorithms and will therefore achieve a solution in the same number of steps.

\vspace{2pt}

This observation is similar to how many advanced distributed training techniques~\citep{huang2019gpipe,zero} are computationally equivalent to regular synchronous training on a single device. For instance, despite using advanced distributed computation strategies, GPipe~\citep{huang2019gpipe} computes exactly the same mathematical expression to obtain gradients and applies those gradients in the same order as any other \textit{synchronous} training algorithm. On the other hand, PipeDream~\citep{pipedream} changes the order in which the updates are applied, introducing the so-called stale gradients~\citep{recht2011hogwild}. This allows PipeDream to improve device utilization but has been shown to reduce the final model quality in some setups~\citep{MLSYS2020_96da2f59}.

\vspace{2pt}

Despite using randomized routing and asynchronous communication between pipeline stages, SWARM still performs optimizer steps synchronously after peers collectively reach the required global batch size (which is a hyperparameter). While different peers may accumulate a different number of samples, they will all use the same gradient after averaging. 
Any peer that fails or does not meet this condition is considered a straggler and must reload its state from neighbors before it can resume training.
This procedure ensures that all surviving peers use non-stale aggregated gradients over the specified batch size when performing the optimizer step. 

\vspace{2pt}

The only deviation from fully synchronous training is that SWARM uses the same approach for CPU offloading as ZeRO-Offload, and by extension, delayed parameter updates (DPU). While DPU was shown not to affect convergence~\citep{zerooffload,stich2020error,arjevani2020tight}, one can disable this functionality and make SWARM fully equivalent to standard training.

\vspace{2pt}

Naturally, these guarantees come at the cost of reduced hardware utilization, as a small portion of devices will need to wait after every step. However, as we show in Section~\ref{sect:experiments_large}, SWARM can still train with competitive training throughput due to the fact that large models are trained with increased batch sizes~\citep{gpt3}.

\section{Additional Details for Section~\ref{sect:experiments_square_cube}}
\label{appendix:detailed_setup}
We benchmark four versions of the Transformer layer:

\begin{itemize}
    \item ``base'': $d_{model} = 768$, $d_{\text{FFN}} = 3072$, 12 heads;
    \item ``xxlarge'': $d_{model} = 4096$, $d_{\text{FFN}} = 16384$, 32 heads;
    \item ``GPT-3''~\citep{gpt3}: $d_{model} = 12288$, $d_{\text{FFN}} = 49152$, 96 heads.
    \item ``Ours'': $d_{model} = 4096$, $d_{\text{FFN}} = 16384$, 32 heads, 3 layers per pipeline stage.
\end{itemize}
\vspace{-4pt}

In Table~\ref{tab:flops_params}, we report FLOP and parameter counts of each version based on the expressions from~\cite{kaplan2020scaling}.
For simplicity, we set up each experiment with 12 Transformer layers using 12 servers (4 for ``Ours'') with a single V100-PCIE GPU each. The servers communicate at 500Mbps under 3--6ms latency. 

\begin{table}[b]
\vspace{-6pt}
\centering
\caption{Parameter and FLOP counts of each architecture.}
\vspace{-4pt}
\label{tab:flops_params}
\begin{tabular}{@{}lcc@{}}
\toprule
Architecture & Parameters & FLOP count \\ \midrule
``base'' & 7.08M & $2.2\times 10^{10}$ \\
``xxlarge'' & 201M & $6.2\times 10^{11}$ \\
``GPT-3'' & 1.81B & $5.5\times 10^{12}$ \\
``Ours'' & 201M & $1.8\times 10^{12}$ \\ \bottomrule
\end{tabular}
\end{table}

Due to a modest communication bandwidth, smaller models spend most of the time waiting for the network. However, that same bandwidth allows for $>80\%$ GPU utilization when dealing with GPT-3-sized layers. If we colocate 3 ``GPT-3'' layers per pipeline stage, the GPU utilization can further improved to $>90\%$.

The time reported in Section~\ref{sect:experiments_square_cube} is the time required to run forward and backward pass for all layers with a batch of 1x512 tokens, not including the Adam updates. All results are averaged over 1000 consecutive batches; the standard deviations are below 0.1\%. All four GPUs are in the same data center but on different servers. Each layer is a \texttt{TransformerEncoderLayer} from PyTorch 1.7.0~\citep{paszke2019pytorch} wrapped with activation checkpointing. We use \texttt{hivemind==0.8.15}~\citep{hivemind_dmoe} with a single synchronous trainer based on the BERT training code from the Transformers library~\citep{wolf-etal-2020-transformers}. However, these results are not specific to hivemind and are likely reproducible in FairScale~\citep{FairScale2021} or PyTorch RPC. The only important detail is that the training code should run as much communication as possible in the background while the GPUs are busy processing batches.
It is important to reuse the same connection for multiple RPC calls so that the TCP buffer does not have to warm up during each call. Also, our implementation performs quantization asynchronously with communication and other computations.

\section{Additional Details for Section~\ref{sect:experiments_large}}
\label{appendix:detailed_large}
\vspace{4pt}

We use the standard Transformer architecture with two modifications: Rotary Positional Embeddings~\citep{su2021roformer} and GeGLU activations~\citep{gated_improve}.
Similarly to other models trained on Pile~\citep{gao2020pile,gptj}, we use the tokenizer of GPT-2~\citep{radford2019language}. Following~\cite{curriculum_minja}, we linearly increase training sequence length during the initial phase. More specifically, we begin training with sequences of up to 256 tokens and increase them to the maximum length of 2048 over the first $12,000$ optimizer steps.
We train the model with LAMB~\citep{lamb}, following the configuration from the original paper for a batch size of 16384. On top of that, we set $\eta=10^{-3}$ and $\beta_2=0.95$ to account for the increased model size.

\section{Additional Scaling Evaluation}\label{appendix:scaling}

In this experiment, we investigate the influence of the number of nodes training with SWARM parallelism on the throughput of the pipeline. Specifically, we measure the performance of training the same model as in Section~\ref{sect:experiments_large} in several configurations that differ in the size of the data-parallel group at each pipeline stage, with the number of single-GPU instances ranging from 8 to 128 (the highest quantity of preemptible nodes that we could reliably maintain for a long time). To isolate the effect of worker heterogeneity, here we use only the T4 accelerators and measure the average performance over 30 minutes of training.

\begin{figure}[b]
    \centering
    \includegraphics[width=0.7\linewidth]{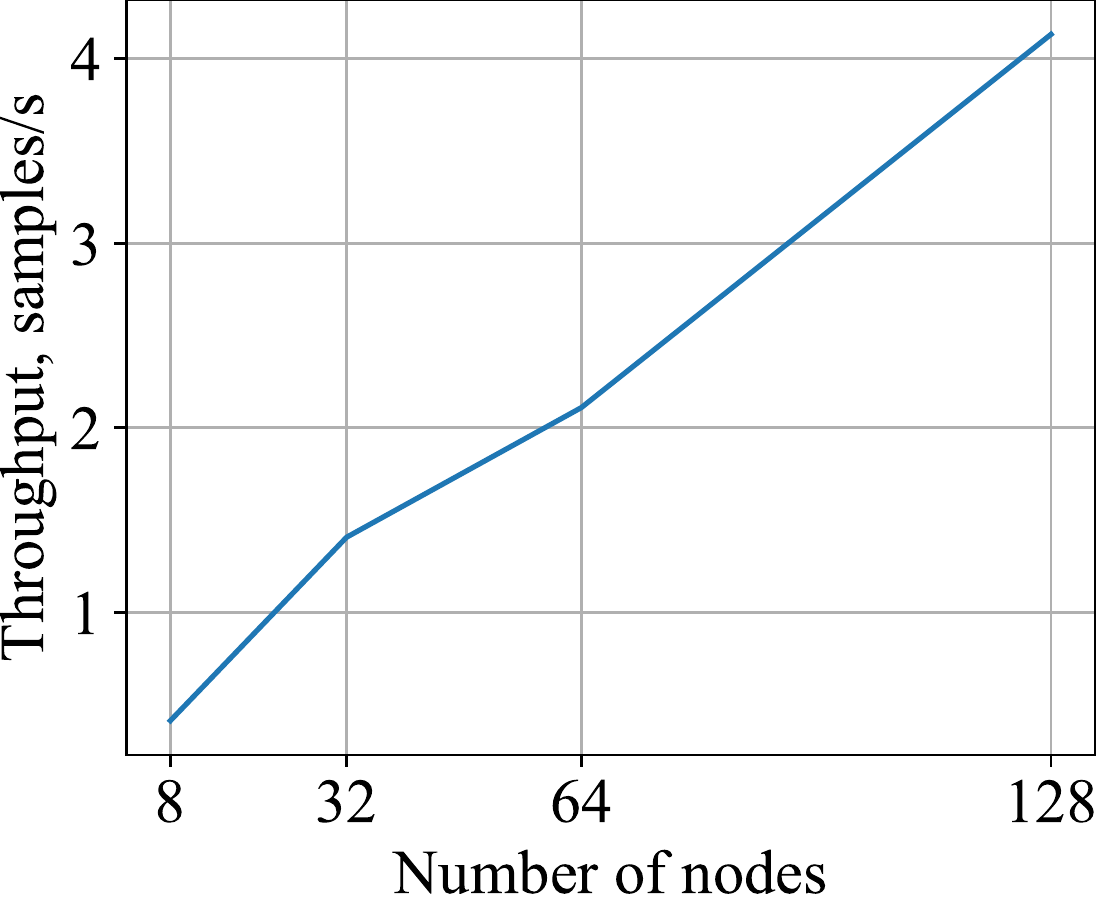}
    \caption{Scaling of SWARM parallelism throughput with the number of nodes.}
    \label{fig:scaling_t4}
\end{figure}

Figure~\ref{fig:scaling_t4} shows the results of our evaluation. It can be seen that the training performance exhibits an approximately linear scaling pattern, which can be explained by the high efficiency of both the stochastic wiring strategy and the auxiliary training components such as the DHT and the All-Reduce protocol used for gradient averaging.

\section{Compression-Aware Architectures}\label{appendix:compression}

Since pipeline parallelism has several distinct points of communication, the network overhead can be reduced considerably by reducing the size of data at these communication points. To exploit this, we develop compression-aware architectures that apply extreme compression at these points. We study two distinct communication bottleneck layers: (1) compression through a linear bottleneck layer, and (2) compression through a bottleneck induced by the maxout activation function~\citep{goodfellow2013maxout}. We also study how compressing the activations and gradients at the communication points to 8 bits affects the predictive performance.

\subsection{Description}\label{appendix:compression_detailed}

\paragraph{Fully connected layers (baseline):} Fully connected layers in models such as Transformers consist of a multilayer perceptron with a single hidden layer and a nonlinear activation function. Without biases and with a residual connection~\citep{resnet} from the inputs to the outputs, this can be described as $\text{MLP}(\mathbf{x}, \mathbf{w}_1, \mathbf{w}_2) = \sigma(\mathbf{x}\mathbf{w}_1)\mathbf{w}_2 + \mathbf{x}$,
where $\mathbf{x}\in\mathbb{R}^{b\times s\times m}$, $\mathbf{w}_1\in\mathbb{R}^{m\times h}$, $\mathbf{w}_2\in\mathbb{R}T^{h\times m}$, and $\sigma(\cdot)$ is a nonlinear activation function such as ReLU~\citep{alexnet}; $b$, $s$, $m$, and $h$ are the batch, sequence, model, and hidden dimensions of the neural network. To compress the output of the MLP layer, we want to apply a compression layer between two consecutive stages. For example, if we have 24 layers and 4 stages, we need 3 compression layers at layers 6, 12, and 18.%

\paragraph{Quantized activations:} A natural way to reduce the communication intensity is to send activations and gradients with respect to activations in reduced precision. However, simply casting tensors to a lower precision may slow down convergence and cause instabilities. Instead, we use dynamic 8-bit quantization with blockwise scaling from~\citep{adam8bit}. This technique reduces communication by ${\approx} 2$x and ${\approx} 4$x for half and full precision, respectively.

On the other hand, quantizing and dequantizing activations can add compute overhead on every microbatch processed. Our implementation circumvents that overhead by performing quantization asynchronously on the CPU. However, this is not required, as blockwise (de)quantization takes less than 1\% of total computation time: see Appendix~\ref{appendix:time_to_solution} for details.

\paragraph{Bottleneck layers:} We experiment with simple bottleneck layers that work by compressing the output features of the MLP by linear projection:
\begin{gather*}
    \text{Bottleneck}(\mathbf{x}, \mathbf{w}_1, \mathbf{w}_2, \mathbf{w}_c, \mathbf{w}_d) = \\ 
    = \text{LayerNorm}(\text{LayerNorm}(\text{MLP}(\mathbf{x}, \mathbf{w}_1, \mathbf{w}_2))\mathbf{w}_c)\mathbf{w_d},
\end{gather*}

where $\mathbf{w}_c\in\mathbb{R}^{m\times c}$, $\mathbf{w}_d\in\mathbb{R}^{c\times m}$ are compression and decompression parameters with compression dimension $c<m$. We find it critical to use layer normalization \cite{ba2016layernorm} to ensure training without divergence. The parameter matrix $\mathbf{w}_c$ resides in one stage and its outputs are transferred to the next stage that holds the parameters $\mathbf{w}_d$, which requires $m/c$ times less communication compared to the original model. Note that adding a bottleneck only adds two linear layers for the forward pass and decreases the size of MLP activations; thus, its computational overhead is negligible (less than 1\% for typical sizes, see Appendix~\ref{appendix:time_to_solution}).

\paragraph{Maxout compression:} Compared to bottleneck compression, maxout compression works by using the maxout activation function~\citep{goodfellow2013maxout} for compression rather than a linear projection. The maxout function of factor $k$ takes inputs with a hidden dimension of $d$ and reduces this dimension by a factor of $k$ by computing the maximum value for each non-overlapping window of $k$ features. We use maxout compression as follows:
\begin{gather*}
    \text{Maxout}(\mathbf{x}, \mathbf{w}_1, \mathbf{w}_2, \mathbf{w}_d) = \\ \text{LayerNorm}(\text{maxout}_k(\text{LayerNorm}(\text{MLP}(\mathbf{x}, \mathbf{w}_1, \mathbf{w}_2))))\mathbf{w_d},
\end{gather*}

where the output is reduced by a factor of $k$ through the maxout function in the previous stage, and then sent to the next stage which holds the decompression matrix $\mathbf{w}_d{\in}\mathbb{R}^{m/k\times m}$.

    \begin{table*}[h!]
    \centering
    \captionof{table}{Performance of compression methods for a Transformer language model with adaptive inputs on WikiText-103. The asterisk denotes that the difference is not statistically significant.}
    \label{tab:steps_to_22}
    \begin{tabular}{@{}lccccc@{}}
    \toprule
    \multirowcell{2}[-0.5ex][l]{Method} & \multirowcell{2}[-0.5ex]{Ppl after\\ 286K steps} & \multirowcell{2}[-0.5ex]{Steps to\\ppl 22} & \multirowcell{2}[-0.5ex]{Data\\ transfer} & \multicolumn{2}{c}{Extra compute} \\ \cmidrule(l){5-6} 
     & &  &  & Absolute & Relative \\ \midrule
    No compression & 21.02 & 1x & 1x & 0 & None \\
    8-bit compression & 21.13 & $\text{0.97x}^{*}$ & 0.5x & 1.2ms & None \tiny{(overlapped)} \\
    Bottleneck & 21.76 & 1.26x & 0.5x & 1.96ms & $\leq 1\%$ \\
    Maxout & 21.83 & 1.28x & 0.5x & 2.04ms & $\leq 1\%$ \\ \bottomrule
    \end{tabular}
    \end{table*}

\subsection{Evaluating the Speed-Quality Tradeoff}
\label{appendix:compression_tradeoff}

While compression techniques reduce the communication overhead, they might also degrade the perplexity reached in a certain time and the final perplexity after a specific number of steps. To study these tradeoffs, we train a Transformer language model with adaptive inputs~\citep{baevski2019adaptiveinputs} on the WikiText-103 dataset and measure how compression-aware architecture variants affect convergence.

Our setup follows that of~\citep{baevski2019adaptiveinputs} with one difference: we use a sequence length of 2048 instead of 3072 to fit this model into our smaller GPUs.
To measure the time to solution, we look at the number of iterations it takes to converge to the training perplexity of \textbf{22}. We evaluate the baseline model and three compression-aware modifications from Section~\ref{appendix:compression_detailed}: bottleneck, maxout, and block-wise dynamic 8-bit quantization, each with 2 pipeline stages and each a compression factor of 2x.

The results can be seen in Table~\ref{tab:steps_to_22}. We can see that 8-bit compression does not degrade the time to 22 perplexity and maintains close to the final perplexity of the baseline. The compression-aware bottleneck and maxout architectures perform equal to each other, but degrade final perplexity slightly and increase time to a perplexity of 22 by 26--28\%.

Using these results, one can determine which method is optimal for their hardware setup. For instance, training with maxout with 2 pipeline stages needs $28\%$ more steps, but accelerates the communication phase by $2$x. If communication is the limiting factor, using maxout or bottleneck compression layers will offer {\it improved} time to perplexity despite the performance degradation. However, the same two techniques would result in slower training in a setup where network bandwidth is unlimited.

In turn, 8-bit quantization reduces communication cost without slowing down per-iteration convergence, making it a ``safe bet'' for situations where the per-iteration convergence must be preserved.
In our large-scale experiments (Section~\ref{sect:experiments_large}), we opt to using quantization since it was enough to fully saturate the GPUs.
If network bandwidth is still a limiting factor, one can combine quantization with bottleneck or maxout compression to further reduce communication.

\vspace{-6pt}
\subsection{Additional Experiments}\label{appendix:compression_extra}

The additional experiments in this section have two purposes: (1) to evaluate how compression methods vary with the number of stages and (2) to evaluate an additional setting that is closer to modern pretraining setups such as GPT-2/3.

While (1) has further implications for scaling, (2) is helpful to account for confounding factors that might have been overlooked in the main experiments on WikiText-103. The WikiText-103 baseline uses non-BPE vocabulary, a long sequence length, and uses adaptive inputs \citep{baevski2019adaptiveinputs}, all of which are not frequently used in modern pretrained Transformers since GPT-2 \citep{radford2019language}.

\begin{table*}
\centering
\caption{Results of language models trained on the OpenWebText Corpus (OWT). The baseline model has 253M parameters and is trained for 8 GPU-days. We apply bottleneck and maxout compression to our baseline in 2 and 4 stages with a compression factor between 2--4x. PTB=Penn Treebank, 1BW=Billion word corpus.}
\label{tab:compression}
\begin{tabular}{lcccccccc}\toprule
                &        &             & \multicolumn{6}{c}{Validation perplexity}                        \\ \cmidrule{4-9} 
Model                &  Stages & Compression & OWT & LAMBADA & WikiText-2 & WikiText-103 & PTB    & 1BW   \\\midrule
Baseline        &   --      & -- &   19.7     &  86.4     &    56.2   &    35.4    &  133.0   &   80.9   \\\midrule
8-bit Quantization  &  2      & 2x        &     19.6       &   89.1     &    {\bf 56.0    }  & {\bf   35.0 }        &  132.7   &  79.8    \\
Bottleneck        &   2      & 2x        &   {\bf19.5  }      &  87.7 &     56.5      &    35.2  &  {129.8 }   &   79.2   \\
Maxout  &  2      & 2x        &     19.6       &  {\bf 85.4 }     &     56.6      &   35.2        &  {\bf 126.8 }    &  {\bf78.8  }   \\\midrule
8-bit Quantization  &  4      & 2x        &    {\bf 19.7  }     & {\bf  87.9  }   &    {\bf 56.3    }  & {\bf   35.2 }        &  {\bf133.9 }  &  {\bf79.8}    \\
Bottleneck         &  4      & 2x       &   21.7          &   100.0      &   66.4   &    40.0   &  149.6      &     89.5  \\
Maxout            &  4      & 2x       &  21.4           &  { 89.9  }    & {  63.9 }      &  {     39.5}      & {142.1   }    &   {86.2 }   \\\midrule
Bottleneck        &   2      & 4x        &   21.6          &    99.8     &   64.8        &   39.6          &  145.6      &   88.3    \\
Maxout            &   2      & 4x        & {\bf20.5}      &  {\bf 89.6 }     &   {\bf  60.0}      & {\bf    37.1 }       & {\bf 141.7 }     &  {\bf83.5}     \\\midrule
Bottleneck      & 4  &  4x   &  28.9  &   141.6      &  100.2 &  58.1  &   235.5     &   118.3    \\
Maxout            &  4      & 4x       &   {\bf 21.3} &   {\bf 93.5 }    &  {\bf 63.6  }      & {\bf 39.2}           &   {\bf 147.7 }   & {\bf 89.1  } \\\bottomrule   
\end{tabular}%
\end{table*}

\vspace{-6pt}
\paragraph{Experimental setup:}

As a baseline, we train a Transformer language model~\citep{transformer} on the OpenWebText corpus~\citep{gokaslan2019openwebtext}. We use the following hyperparameters: sequence size 512, 16 layers with model dimension 1024, and hidden dimension 4096 for a total of 253M parameters. We use byte pair encoding~\citep{sennrich-etal-2016-neural,radford2019language} with a vocabulary size of 50264 symbols. We do not use dropout or other regularization, since our models underfit. We run these experiments in Fairseq~\citep{Ott2019fairseqAF}.

We test bottleneck and maxout compression for a compression factor of 50\% and 75\% compared to the original size over two and four stages. We look at how using these compression-aware architectures affects the performance compared to the compression that they achieve.
\vspace{-6pt}

\paragraph{Results:} The results of our compression-aware architectures are shown in Table~\ref{tab:compression}. We can see that while the bottleneck architecture is competitive with maxout for a compression factor of 2x with two stages, maxout has better perplexities if more stages or a higher compression ratio is used. The out-of-distribution perplexities vary consistently with the in-distribution perplexity, which suggests compression-aware architectures do not degrade the out-of-distribution performance more than the in-distribution performance. As such, the maxout compression is an effective technique to reduce the bandwidth requirements of pipeline parallel training further.

While the 8-bit blockwise quantization can only compress the activations by a factor of two (16-bit $\rightarrow$ 8-bit), it does not affect the quality as much when compared to the baseline. As such, the 8-bit quantization appears to be a reliable default choice to reduce the communication overhead for pipeline parallelism.

When considered together with the square-cube law for distributed training and SWARM parallelism, compression-aware architectures allow for better scaling of large neural networks trained over preemptible low-bandwidth peers. Thus, compression-aware architectures improve the accessibility and affordability of training large models outside HPC environments.

\vspace{-6pt}

\section{Time To Solution}\label{appendix:time_to_solution}

In this section, we evaluate the compression-aware techniques proposed in Appendix~\ref{appendix:compression_detailed} from a practitioner's point of view. A natural way to compare these techniques is in terms of ``the time to solution'', i.e., the wall-clock time it takes to achieve the desired validation objective.
In practice, this time depends on three main factors: the compression strategy, the distributed training algorithm, and the computational infrastructure.

In order to disentangle these factors, we first address the relationship between the training algorithm and the infrastructure.
As we discuss in Section~\ref{sect:method_swarm} (and later in Appendix~\ref{appendix:equivalence}), SWARM parallelism has the same per-iteration behavior as other synchronous methods. Theoretically, the choice of an optimal training system should come down to whichever algorithm has the highest training throughput.

To verify this argument in practice, we compare the per-iteration and per-hour performance of SWARM against fully synchronous training. For this experiment, we train the ALBERT model~\citep{albert} on the WikiText-103 dataset~\citep{wikitext103}. We use the ALBERT-Large architecture with 4 layer groups that correspond to 4 SWARM stages \textit{without the architecture modifications from Appendix~\ref{appendix:compression}}. We follow the exact hyperparameters from the original paper: for example, we use the LAMB optimizer~\citep{lamb} with the batch size of 4096 and the sequence length of 512. We train this model in three setups: traditional distributed training with 8 V100 workers, SWARM with 8 preemptible V100 GPUs, and SWARM with 32 preemptible T4 workers.

\begin{table}[t]
\centering
\captionof{table}{Training time and costs.}
\label{tab:cost}
\begin{tabular}{@{}lccc@{}}
\toprule
\multirowcell{2}[-0.5ex][l]{Setup} & \multirowcell{2}[-0.5ex]{Time, hours} & \multicolumn{2}{c}{Cost, \$} \\
\cmidrule(lr){3-4} 
    & & Hourly & Total \\
\midrule
$8\times V100$, reliable         &175.4&7.834&1374\\
\midrule
$8\times V100$, preemptible           &192.6&5.383&1037\\
\midrule
$32 \times T4$, preemptible           &140.8&3.536&497.8\\
\bottomrule
\end{tabular}
\end{table}

To quantify the time to solution, we measure the wall time required to achieve the ALBERT objective equal to \textbf{1.5}. Additionally, we report the per-hour cost of each experimental setup and the total cost of achieving a loss of 1.5 using public cloud provider pricing estimates in \cref{tab:cost}.

\begin{figure}
\centering
\includegraphics[width=\linewidth]{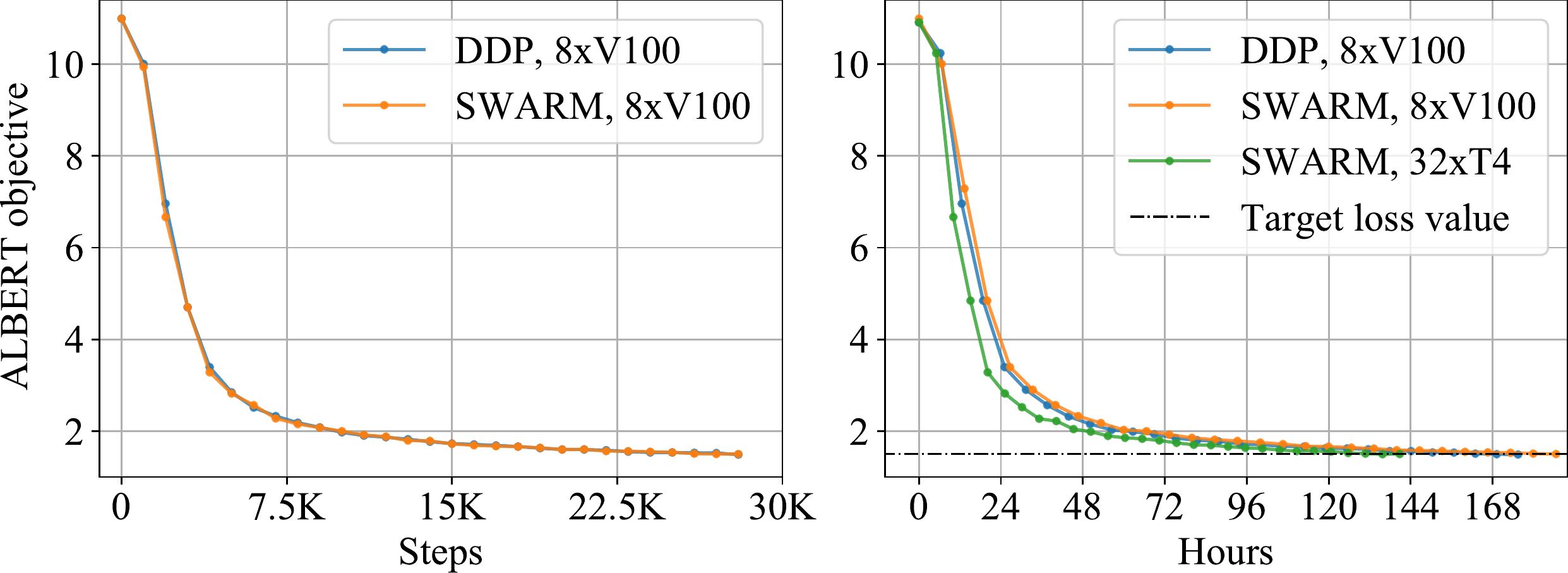}
\captionof{figure}{Convergence curves of ALBERT with SWARM and standard data-parallel training.}
\label{fig:convergence_iterations}
\end{figure}

Figure~\ref{fig:convergence_iterations} demonstrates that SWARM matches the per-iteration learning curves of traditional distributed training (PyTorch DistributedDataParallel) up to the variation comparable to caused by changing the random seed. However, SWARM parallelism can achieve the loss of 1.5 more cost-efficiently and faster by using preemptible instances. In turn, \textit{when forced to use homogeneous and reliable GPUs}, SWARM would have slightly inferior performance compared to conventional algorithms, which was first demonstrated in Section~\ref{appendix:training_throughput}.

\end{document}